\documentclass[AMA,Times1COL]{WileyNJDv5} 

\usepackage{amsmath,amsfonts,amssymb}
\usepackage{subfigure,graphicx}

\articletype{Article Type}%
\received{Date Month Year}
\revised{Date Month Year}
\accepted{Date Month Year}
\journal{arXiv}
\volume{00}
\copyyear{2025}
\startpage{1}

\DeclareMathOperator{\cov}{cov}
\DeclareMathOperator{\E}{E}
\def\half{{\textstyle\frac{1}{2}}}
\let \Pr \relax
\DeclareMathOperator{\Pr}{P}
\DeclareMathOperator{\rd}{d}
\DeclareMathOperator{\rk}{rank}
\def\Rset{\mathbb{R}}
\DeclareMathOperator{\SE}{SE}
\DeclareMathOperator{\var}{var}
\DeclareMathOperator{\ovec}{vec}
\DeclareMathOperator{\vech}{vech}

\def\github{\textsf{github}}

\def\R{\textsf{R}}

\newcommand{\bs}[1]{\mbox{\boldmath $#1$}} 
\newcommand{\what}[1]{\widehat{#1}}

\DeclareGraphicsExtensions{.pdf,.jpg,.png}

\begin{document}

\title{A new coefficient to measure agreement between continuous variables}

\author[1]{Ronny Vallejos}
\author[1]{Felipe Osorio}
\author[1]{Clemente Ferrer}

\authormark{Vallejos, Osorio and Ferrer}
\titlemark{A new measure of agreement}

\address[1]{\orgdiv{Departamento de Matem\'atica}, \orgname{Universidad T\'ecnica Federico Santa Mar\'ia},
  \orgaddress{\state{Valpara\'iso}, \country{Chile}}}

\corres{Ronny Vallejos, Departamento de Matem\'atica, Universidad T\'ecnica Federico Santa Mar\'ia, 
  Avenida Espa\~na 1680, Valpara\'iso, Chile. \email{ronny.vallejos@usm.cl}}



\abstract[Abstract]{Assessing agreement between two instruments is crucial in clinical studies to 
evaluate the similarity between two methods measuring the same subjects. This paper introduces a 
novel coefficient, termed $\rho_1$, to measure agreement between continuous variables, focusing on 
scenarios where two instruments measure experimental units in a study. Unlike existing coefficients, 
$\rho_1$ is based on $L_1$ distances, making it robust to outliers and not relying on nuisance parameters. 
The coefficient is derived for bivariate normal and elliptically contoured distributions, showcasing its 
versatility. In the case of normal distributions, $\rho_1$ is linked to Lin's coefficient, providing a 
useful alternative. The paper includes theoretical properties, an inference framework, and numerical 
experiments to validate the performance of $\rho_1$. This novel coefficient presents a valuable tool 
for researchers assessing agreement between continuous variables in various fields, including clinical 
studies and spatial analysis.}



\maketitle

\renewcommand\thefootnote{}
\footnotetext{\textbf{Abbreviations:} CCC, concordance correlation coefficient; EC, elliptically contoured.}

\renewcommand\thefootnote{\fnsymbol{footnote}}
\setcounter{footnote}{1}

\section{Introduction and Motivation}\label{sec:intro}

The study of agreement between two variables has become relevant during the last decades in a number of 
different fields, and often arises in research. An example of this happens when two instruments measure 
experimental units in a study, and it is of interest to determine how well these instruments agree. If 
there exist a gold standard the problem  is reduced to  decide whether the approximate method can act as 
a reasonable replacement for the gold standard \citep{Laurent:1998}.

For data measured on a continuous scale, several approaches have been defined to assess the agreement of 
the outcomes. The Pearson correlation coefficient, the pair \emph{t}-test,  the coefficient of variation, 
among others. However, none of these coefficients was created to assess the agreement between two sets of 
measurements. \cite{Lin:1989} developed a concordance correlation coefficient (CCC), which is based on the 
expected squared difference between two variables and measures the deviation of observations from the 45 
degree straight line passing through the origin. This coefficient inherits several properties of Pearson's 
correlation. It is bounded in absolute value by 1, and has an appealing interpretation as the product of 
an accuracy coefficient that measures the agreement between two marginal distributions and another coefficient 
that measures the association or correlation between the measurements \citep{Guo:2007}. There have been 
some extensions of this CCC that use several measuring instruments and techniques to evaluate the agreement 
between two instruments; these efforts have led to interesting graphical tools \citep{Hiriote:2011, Stevens:2017}. 
In the context of goodness of fit, \cite{Vonesh:1996} proposed a modified Lin's CCC for choosing models that 
have a better agreement between the observed and the predicted values. Additionally, \cite{Vallejos:2020} 
extended Lin's coefficient for two continuous spatial variables that are measured at the same locations in 
space. Recently, \cite{Stevens:2017} and \cite{Chodhary:2017} developed the probability of agreement, and 
\cite{Leal:2019} studied the local influence of the CCC and the probability of agreement considering both 
first- and second-order measures under the case-weight perturbation scheme. Lately, \cite{Bottai:2022} examined 
a linear predictor coefficient that maximizes Lin's coefficient. Finite sample and asymptotic properties 
leading to confidence intervals were studied by \cite{Kim:2023}.

To characterize the CCC coefficient, assume that $(X_1,X_2)^\top\sim \mathsf{N}_2(\bs{\mu},\bs{\Sigma})$, where 
$\bs{\mu} = (\mu_1,\mu_2)^\top$ denotes the mean vector, and 
\[ 
  \bs{\Sigma} = \begin{pmatrix}
    \sigma_{11} & \sigma_{12} \\
    \sigma_{21} & \sigma_{22}
  \end{pmatrix},
\] 
being a covariance matrix. Lin's coefficient of concordance is defined through
\begin{equation}\label{eq:Lin}
  \rho_c = 1 - \frac{\E\{(X_1 - X_2)^2\}}{\E\{(X_1 - X_2)^2 : \sigma_{12} = 0\}} = \frac{2\sigma_{12}}
  {\sigma_{11} + \sigma_{22} + (\mu_1 - \mu_2)^2},
\end{equation}
and characterize the deviation of the variables from the straight line with zero slope, through the term 
$\E\{(X_1 - X_2)^2\}$, properly normalized to have a coefficient bounded in absolute value by 1.

The inference of $\rho_c$ was carried out substituting the sample moments of an independent bivariate sample 
to obtain $\what{\rho}_c$. Precisely, 
\begin{equation}\label{eq:est-Lin}
  \what{\rho}_c = \frac{2s_{12}}{s_{11} + s_{22} + (\overline{x}_1 - \overline{x}_2)^2},
\end{equation}
where $s_{12}$ is the sample covariance between $X_1$ and $X_2$, $s_{11}$ and $s_{22}$ are the sample variances 
of $X_1$ and $X_2$ respectively, and similarly $\overline{x}_1$ and $\overline{x}_2$ are the sample means of $X_1$ 
and $X_2$. The Fisher $Z$-transformation 
\[
  \what{Z} = \tanh^{-1}(\what{\rho_c}) = \frac{1}{2}\ln\Big(\frac{1 + \what{\rho}_c}{1 - \what{\rho}_c}\Big),
\]
was used to demonstrate the asymptotic normality of $\what{Z}$. Thus, $\sqrt{n}(\what{Z} - Z) \stackrel{D}{\rightarrow} 
\mathsf{N}(0, v^2)$, as  $n\to\infty$, where $Z = \tanh^{-1}(\rho_c) = \frac{1}{2}\ln\big(\frac{1 + \rho_c}{1 - \rho_c}\big)$, 
and (see the Appendix of \cite{Lin:1989} and \cite{Lin:2000})
\begin{equation}\label{eq:var-Lin}
  v^2 = \frac{1}{n-2}\Big\{\frac{(1 - \rho^2)\rho_c^2}{(1 - \rho_c^2)\rho^2} + \frac{2\rho_c^3(1 - \rho_c)u^2}
  {\rho(1 - \rho_c^2)^2} -\frac{\rho_c^4u^4}{2\rho^2(1 - \rho_c^2)^2}\Big\},
\end{equation}
with $u = (\mu_1 - \mu_2)/\sqrt{\sigma_{11}\sigma_{22}}$ and $\rho = \sigma_{12}/\sqrt{\sigma_{11}\sigma_{22}}$
being the Pearson correlation coefficient. As a consequence, the limiting distribution of $\what{\rho}_c = \tanh(\what{Z})$ 
is normal with mean $\rho_c$ and variance $(1 - \rho_c^2)^2 v^2/n$. Using this result, approximate confidence intervals 
and hypothesis testing can be constructed for $\rho_c$.

\begin{figure}[!ht]
  \vskip -1.75em
  \centering
  \subfigure[]{
    \includegraphics[width = 0.45\linewidth]{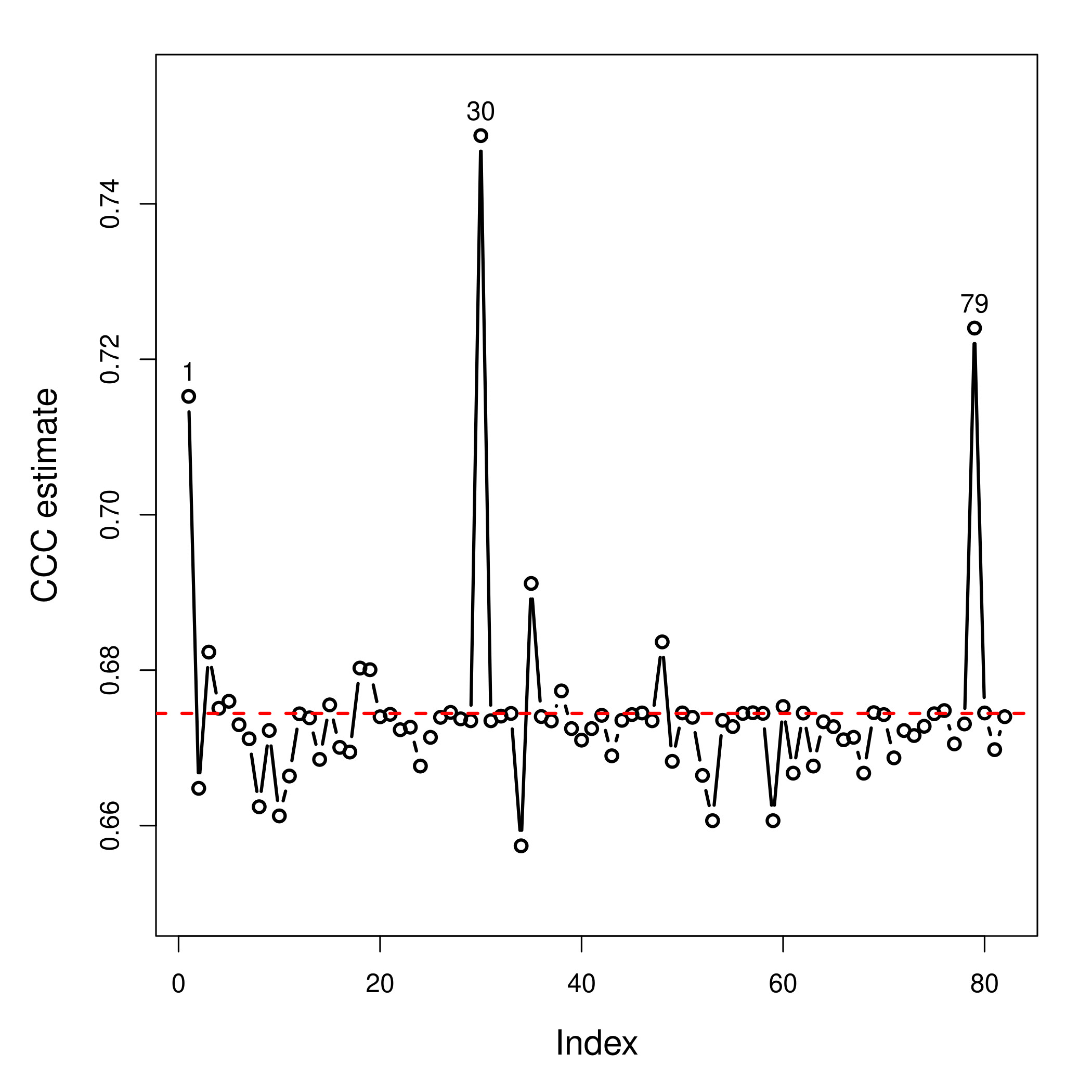}
  }
  \subfigure[]{
    \includegraphics[width = 0.45\linewidth]{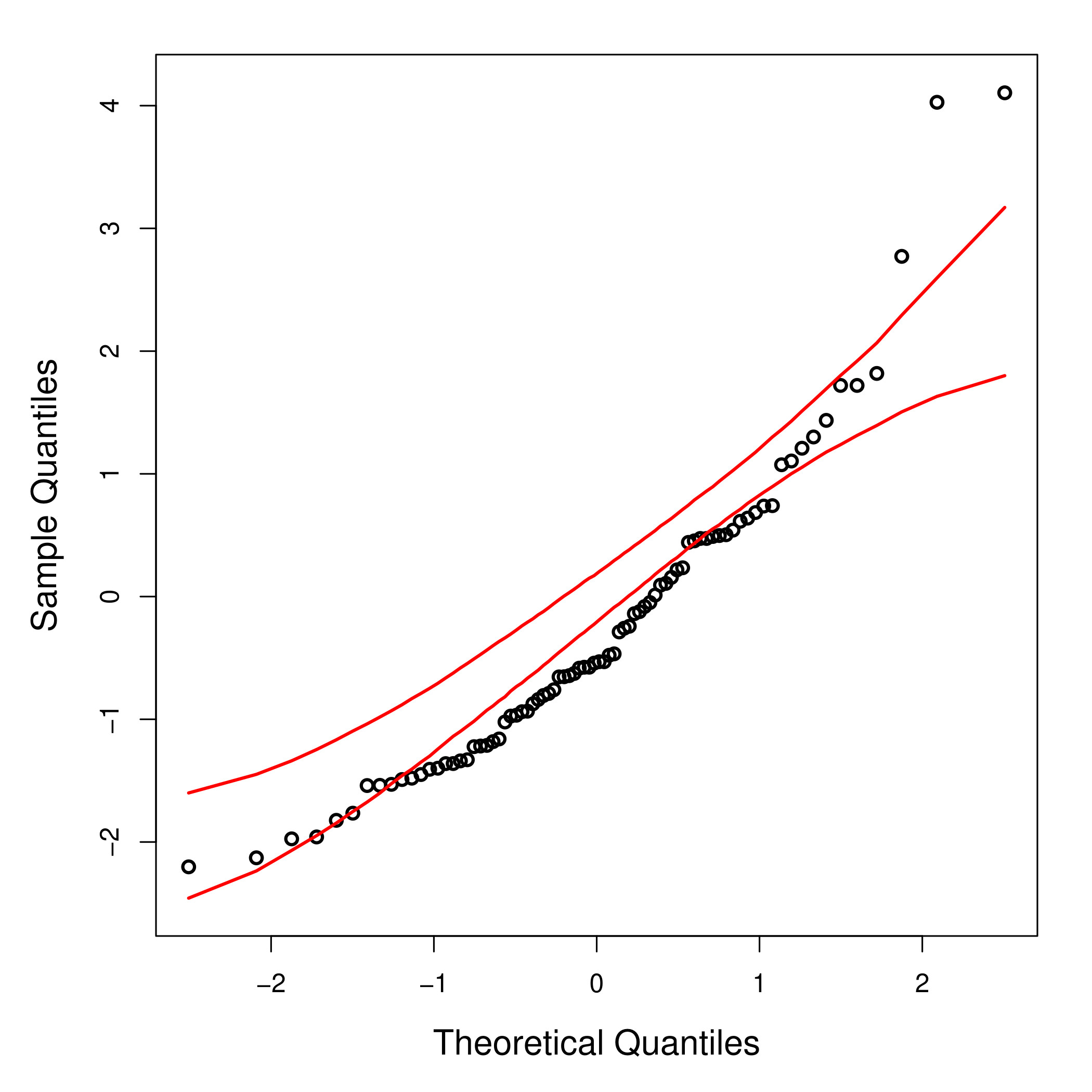}
  }
  \caption{(a) Case deletion plot: The effect on the CCC estimates when the $i$th observation is removed from the dataset; 
  (b) QQ-plot of the transformed Mahalanobis distances associated with variables $X_1$ and $X_2$.}\label{fig:intro}
\end{figure}

As an illustration, consider the clinical study conducted by \cite{Svetnik:2007} to compare the automated and 
semi-automated scoring Polysomnographic (PSG) recordings used to diagnose transient sleep disorders. The analysis 
considered 82 patients who were administrated 10 mg of a sleep inducing drug called Zolpidem. The study recorded 
measurements of latency to persistent sleep (LPS). The observations were obtained using six different methods. 
In this work, following the work of \cite{Leal:2019}, we focus on two of the six methods: fully manual scoring 
(Manual) and automated scoring by the Morpheus software (Automatic). Let $X_1$ and $X_2$ be the $\log(\mathrm{LPS})$ 
variables with the manual and automatic methods respectively. Also assume that $(X_1,X_2)^\top\sim \mathsf{N}_2(\bs{\mu}, 
\bs{\Sigma})$. Then the estimated CCC is given by $\what{\rho}_c = 0.675$. This value is in agreement with the 
0.650 cut-off point proposed by \cite{McBride:2005}, who suggested a low degree of agreement between these two 
variables. One of the possible reasons to have a low degree of agreement is the fact that the data contains outliers 
or influential points that affect the relationship between $X_1$ and $X_2$, and also impact the overall fitting of 
the observations  to a bivariate normal distribution. To gain more insights in this respect, we display two plots 
in Figure \ref{fig:intro}. The plot on the left shows the CCC estimate when the $i$th observation is deleted. 
There is a clear effect of observations 30 and 79 on the estimations. Figure \ref{fig:intro}\,(b) shows the impact 
of theses observations on the normality assumption. We summarize this brief data analysis saying that there are at 
least two aspects that need to be consider in the presence of influential observations: the robustness of the coefficient, 
and the misspecification of the distribution. In this context, \cite{King:2001} studied a robust version of Lin's 
coefficient applying an alternative distance function to compute the CCC index. Subsequently, \cite{Tashakor:2019} 
introduced an alternative  CCC, that is a particular case of the robust estimator studied in \cite{King:2001}, but 
yields robust $L$-statistics. On the other hand, the effects of non-normality on the distribution of the sample 
product moment correlation coefficient were investigated in \cite{Devlin:1975} among others through some simulation 
studies.

In this paper, we introduce and propose a new robust estimator of CCC that is based on $L_p$ distances and does not 
depend on nuisance parameters that need to be estimated. Our proposal can be computed for bivariate normal random 
vectors, but is it not limited solely to Gaussianity. Indeed, in this work our coefficient and Lin's coefficient are 
also derived for elliptically contoured (EC) bivariate distributions. under certain conditions a relationship between 
these two coefficients is also  established. besides the construction of the new index, we address the estimation and 
obtain the asymptotic variance of the estimation. In addition some discussion provides a criterion to select $p$ in 
an optimal way. Numerical experiments are carried out to gain more insights into the performance of the coefficient. 
Experiments with real data are also provided to enhance the presentation. The proofs of the main results are relegated 
to an Appendix.

\section{A new coefficient}

\subsection{Construction}

There are several ways to measure the discrepancy between variables $X_1$ and $X_2$. The underlying distance function 
associated with the CCC index is $\E\{\varphi(X_1 - X_2)\}$, for $\varphi(z) = z^2$. \cite{King:2001} provided a 
suitable function $\varphi(\cdot)$ to robustify original Lin's proposal so that the new coefficient is reduced to 
Lin's coefficient when $\varphi(z) = z^2$. In order to try different $\varphi(\cdot)$ functions in their study, Huber's 
function, as well as the absolute distance function of the form $\varphi(z) = \lvert z\rvert^\delta$, where $1 < \delta \leq 2$, 
were considered. As a result of the numerical experiments, the best performance of the robust estimators was obtained 
for the Winzorized function
\[
  \varphi(z) = \begin{cases} 
    \lvert z\rvert^\delta, & \lvert z\rvert\leq z_0, \\
    \lvert z_0\rvert^\delta, & \lvert z\rvert > z_0,
  \end{cases}
\]
for $\delta = 1$ and $\delta = 1.5$, and $z_0 > 0$, when the underlying data contain contamination. We emphasize 
that the approach studied by \cite{King:2001} consist of an estimator that is constructed using a numerator based 
on a difference between two expected values of a function $\varphi(X_1 - X_2)$. In this paper to look at the agreement 
between two continuous variables we consider a function $\varphi(z) = \lvert z\rvert$ to define
\begin{equation}\label{eq:L1}
  \rho_1 = 1 - \frac{\E(\lvert X_1 - X_2\rvert)}{\E(\lvert X_1 - X_2\rvert: \sigma_{12} = 0)}.
\end{equation}
Coefficient \eqref{eq:L1} can be computed without the need of estimating nuisance parameters as usually happens in 
robustness. As in \eqref{eq:Lin}, if we assume that $(X_1, X_2)^\top\sim \mathsf{N}_2(\bs{\mu},\bs{\Sigma})$, then 
\begin{equation}\label{eq:CCC1}
  \rho_1 = 1 -\frac{\gamma\big(1 - 2\Phi(-\gamma/\tau)\big) + \tau\sqrt{2/\pi}\exp\big(-\half(\gamma/\tau)^2\big)}
  {\gamma\big(1 - 2\Phi(-\gamma/\sqrt{\sigma_{11} + \sigma_{22}})\big) + \sqrt{2(\sigma_{11} + \sigma_{22})/\pi}
  \exp\big(-\half\gamma^2/(\sigma_{11} + \sigma_{22})\big)},
\end{equation}
where $\gamma = \mu_1 - \mu_2$, $\tau^2 = \sigma_{11} + \sigma_{22} - 2\sigma_{12}$ and $\Phi(\cdot)$ is the cumulative 
distribution function of the standard normal distribution. In particular, if $\mu_1 = \mu_2$, the coefficient \eqref{eq:CCC1} 
can be expressed as
\begin{equation}\label{eq:coef}
  \rho_1 = 1 - \sqrt{\frac{\sigma_{11} + \sigma_{22} - 2\sigma_{12}}{\sigma_{11} + \sigma_{22}}}.
\end{equation}
To elucidate the equality of the means, we can consider a test of the form $H_0: \mu_1 = \mu_2$. When the hypothesis 
$H_0$ is rejected, computing $\rho_1$ from \eqref{eq:CCC1} is straightforward since there is good software for evaluating $\Phi(\cdot)$.

\subsection{Properties}

Now, to explore how the coefficients \eqref{eq:Lin} and \eqref{eq:L1} look like for an elliptically contoured distribution, 
consider the following result (see page 43 of \cite{Fang:1990}). Suppose that $\bs{X}$ is a $k$-variate random vector such 
that $\bs{X}\sim\mathsf{EC}_k(\bs{\mu},\bs{\Sigma};\psi)$ where $\psi$ is a function of scalar variable, called the 
characteristic generator of the random vector $\bs{X}$, $\rk(\bs{\Sigma}) = k$ and $R^2 \stackrel{d}{=} (\bs{x} - \bs{\mu})^\top 
\bs{\Sigma}^{-1}(\bs{x} - \bs{\mu})$ is known as radial random variable. Then
\[
  \E(\bs{X}) = \bs{\mu}, \qquad \cov(\bs{X}) = \frac{\E(R^2)}{k}\bs{\Sigma},
\]
provided that $\E(R^2) < \infty$. If $(X_1,X_2)^\top\sim \mathsf{EC}_2(\bs{\mu},\bs{\Sigma};\psi)$. Then, 
$Z = X_1 - X_2 \sim \mathsf{EC}_1(\gamma, \tau^2; \psi)$. Assuming that $\E(R^2) < \infty$, we have that 
$\E(Z) = \gamma$, and $\var(Z) = \E(R^2)\tau^2/2$. Thus $\E(Z^2) = \E(R^2)\tau^2/2 + \gamma^2$. If $X_1$ 
and $X_2$ are no correlated, it follows that $\E(Z^2:\sigma_{12} = 0) = \E(R^2)(\sigma_{11} + \sigma_{22})/2 
+ \gamma^2$. Then Lin's coefficient for a bivariate EC distribution is
\begin{align}
  \rho_c & = 1 - \frac{\E(Z^2)}{\E(Z^2: \sigma_{12} = 0)} = 1 - \frac{\gamma^2 + \E(R^2)\tau^2/2}{\gamma^2 
  + \E(R^2)(\sigma_{11} + \sigma_{22})/2} = \frac{\E(R^2)\sigma_{12}}{\gamma^2 + \E(R^2)(\sigma_{11} + \sigma_{22})/2} \nonumber \\ 
  & = \rho\,C_{12}(R^2), \label{eq:rhoc-EC}
\end{align}
where $\rho = \sigma_{12}/\sqrt{\sigma_{11}\sigma_{22}}$, and $C_{12}(R^2) = 2[b + b^{-1} + a^2/(\E(R^2)/2)]^{-1}$,
with $a = \gamma/(\sigma_{11}\sigma_{22})^{1/4}$, and $b = \sqrt{\sigma_{11}/\sigma_{22}}$. It is easy to see that
for $(X_1,X_2)^\top \sim \mathsf{N}_2(\bs{\mu},\bs{\Sigma})$ we have $R^2$ follows a chi-squared distribution, thus 
$\E(R^2) = 2$, and we recovered the definition of CCC given by Lin.

In the sequel, we will assume that the elliptical contoured distributions have density. In which case we 
write $\bs{X}\sim\mathsf{EC}_k(\bs{\mu},\bs{\Sigma};g)$, with probability density function, given by 
\[
  f(\bs{x};\bs{\mu},\bs{\Sigma}) = C_g \lvert\bs{\Sigma}\rvert^{-1/2} g[(\bs{x}-\bs{\mu})^\top
  \bs{\Sigma}^{-1}(\bs{x}-\bs{\mu})],
\]
where $g$ is known as the density generator function\cite{Fang:1990} and $C_g$ denotes a constant of integration. 
Consider the following result.

\begin{proposition}\label{prop:1}
  Assume that $(X_1,X_2)^\top \sim \mathsf{EC}_2(\bs{\mu},\bs{\Sigma};g)$. Then 
  \begin{equation}\label{eq:CCC1-EC}
    \rho_1 = 1 - \frac{\displaystyle\gamma\Big(1 - 2C_g\int_{-\infty}^{-\alpha_1}g(r^2)\rd r\Big) - 2C_g\tau\int_{-\infty}^{-\alpha_1} 
    r g(r^2)\rd r}{\displaystyle\gamma\Big(1 - 2C_g\int_{-\infty}^{-\alpha_2}g(r^2)\rd r\Big) - 2C_g\sqrt{\sigma_{11} 
    + \sigma_{22}}\int_{-\infty}^{-\alpha_2} r g(r^2)\rd r},
  \end{equation}
  where $\alpha_1 = \gamma/\tau$ and $\alpha_2 = \gamma/\sqrt{\sigma_{11}+\sigma_{22}}$.
\end{proposition}

In the simple case when $\mu_1 = \mu_2$, it is straightforward to see from \eqref{eq:coef} that
\begin{equation}\label{eq:relationship}
  \rho_1 = 1 - \sqrt{\frac{\sigma_{11} + \sigma_{22} - 2\sigma_{12}}{\sigma_{11} + \sigma_{22}}}
  = 1 - \sqrt{1 - \rho_c}.
\end{equation}
Alternatively, in this case $Z = X_1 - X_2\sim \mathsf{EC}_1(0,\tau^2;g)$ and the pdf of $Z$ is given by
\begin{equation}\label{eq:density}
  f(z) = \frac{C_g}{\sqrt{\sigma_{11} + \sigma_{22} - 2\sigma_{12}}}\,g\Big(\frac{z^2}{\sigma_{11} + \sigma_{22} 
  - 2\sigma_{12}}\Big), \qquad z\in\Rset.
\end{equation}
Then the expected value $\E(\lvert Z\rvert)$ can be obtained via integration.

\begin{figure}[!ht]
  \vskip -1.75em
  \centering
  \subfigure[]{
    \includegraphics[width = 0.45\linewidth]{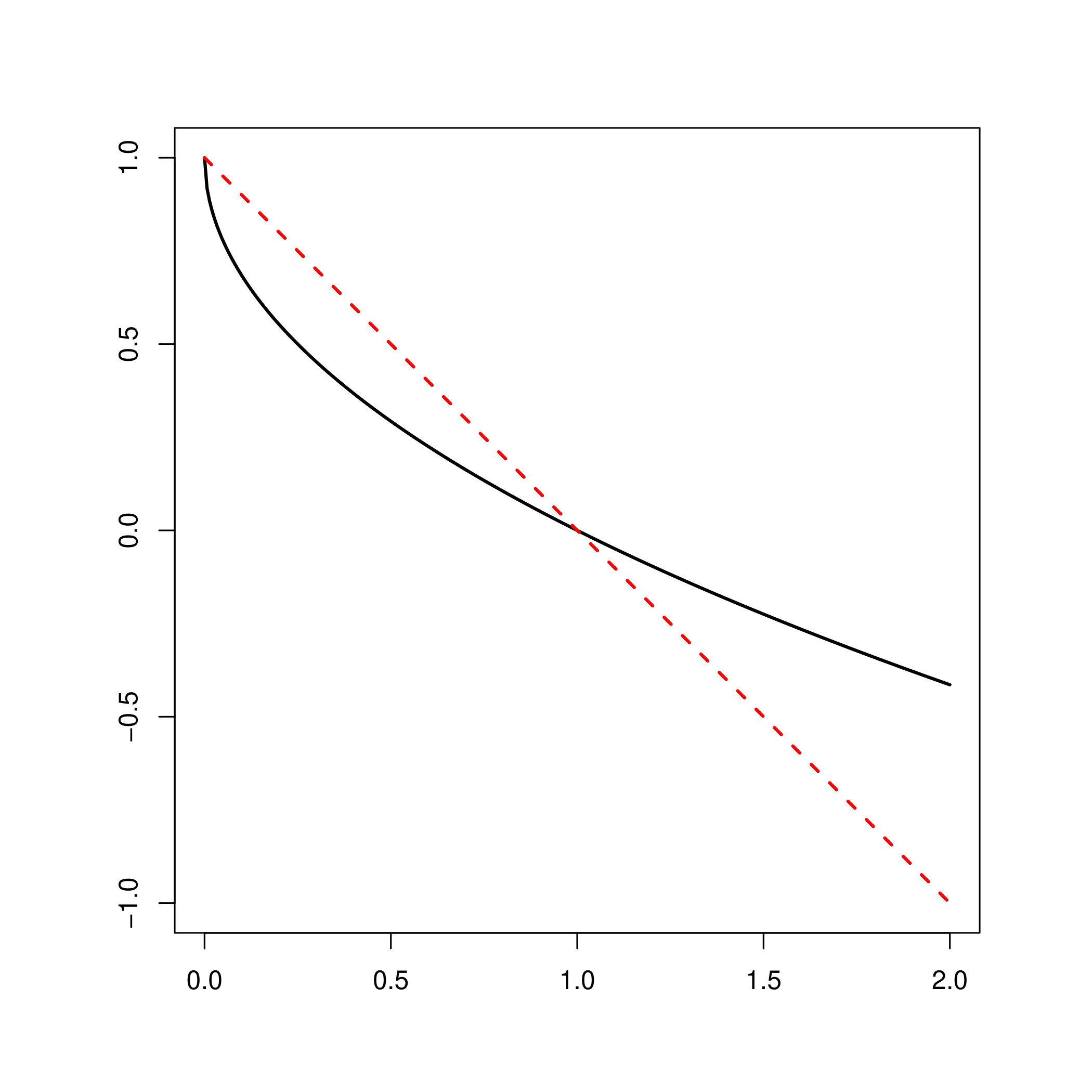}
  }
  \subfigure[]{
    \includegraphics[width = 0.45\linewidth]{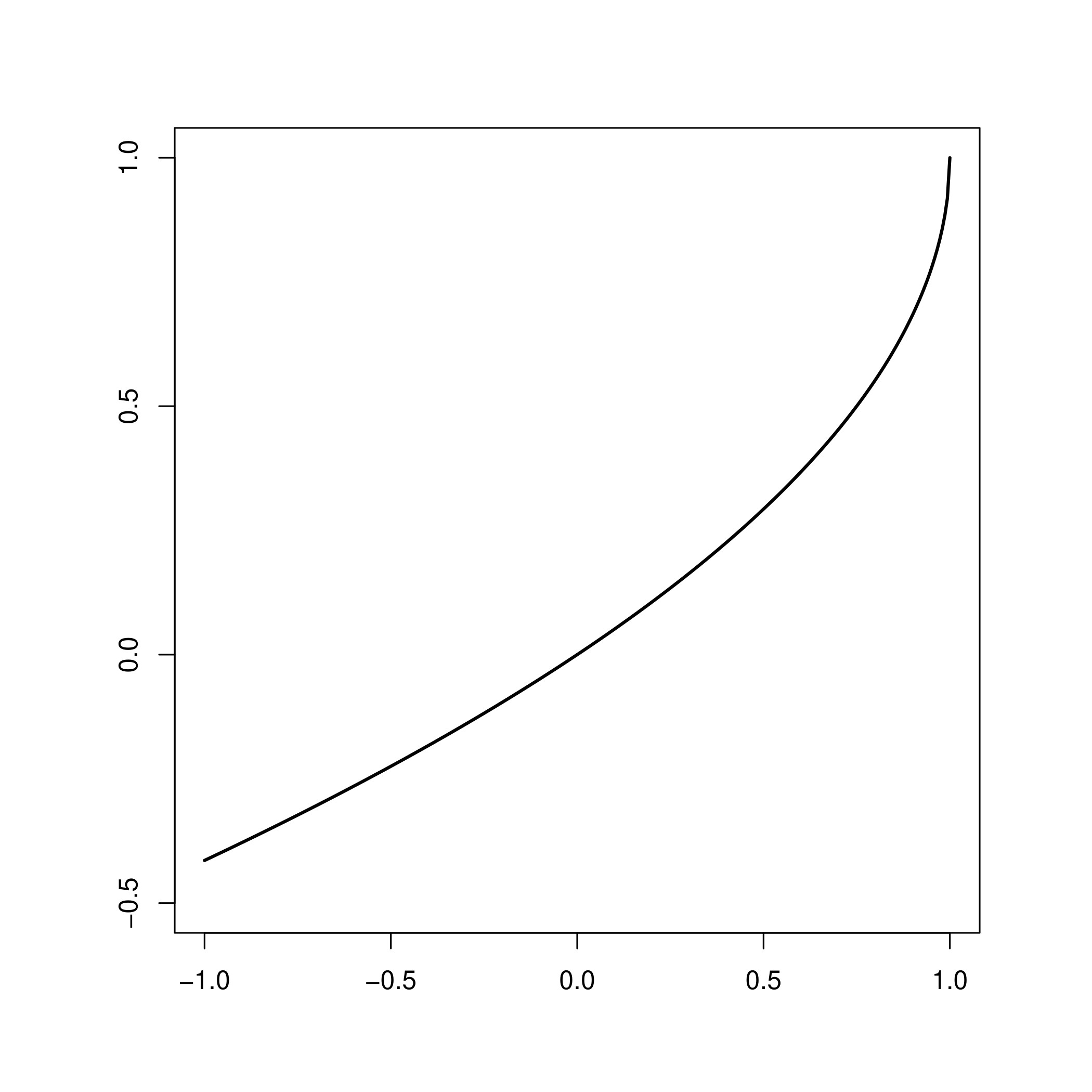}
  }
  \caption{(a) plot of $\rho_c$ (dashed line) and $\rho_1$ (solid line) versus $\xi$, for $\xi \in [0,2]$; 
  (b) plot of $\rho_1$ versus $\rho_c$.}\label{fig:compare}
\end{figure}

We should note that in this case, $\rho_1$ corresponds to \eqref{eq:coef}, which was derived assuming normality. 
This implies that these measures of agreement rely on the variances and covariance, regardless of the means of 
variables $X_1$ and $X_2$, which are not considered in measuring the discrepancy between them. Additionally, it 
is surprising that $\rho_1$ does not depend on $g(\cdot)$. Moreover, when $\mu_1 = \mu_2$, $\rho_c = 1 - (\sigma_{11}
+ \sigma_{22} - 2\sigma_{12})/(\sigma_{11} + \sigma_{22})$. By defining $\xi := (\sigma_{11} + \sigma_{22} 
- 2\sigma_{12})/(\sigma_{11}+\sigma_{22})$, we can compare the functions $\xi\mapsto 1-\sqrt{\xi}$ and $\xi\mapsto 
1-\xi$ within a bounded interval since $\rho_c\in[-1,1]$. To illustrate this, we have plotted $\rho_c$ and $\rho_1$ 
against $\xi$ for $\xi \in [0,2]$ in Figure \ref{fig:compare}\,(a).

We distinguish three interesting cases, i) If $\xi\in(0,1]$, then $\rho_c > \rho_1$. ii) If $\xi=0$, then $\rho_c
= \rho_1$. iii) If $\xi\in[-1,0)$, then $\rho_c < \rho_1$. Additionally, we observe that the minimum value for 
$\rho_1$ is approximately $1 - \sqrt{2}\approx 0.41$. These patterns are visually represented in Figure \ref{fig:compare}\,(b), 
where Equation \eqref{eq:relationship} has been plotted.

Due to the fact that the minimum value of $\rho_1$ is not -1, this coefficient can effectively measure positive 
agreement between $X_1$ and $X_2$. This characteristic is not uncommon, as there are numerous practical scenarios 
where quantifying spatial agreement between two georeferenced variables is important. In such cases, the association 
between the variables relies on the distance between pairs of points in space (see, for instance \cite{Vallejos:2020}). 
An alternative method for defining scaled coefficients, such as \eqref{eq:coef}, within the interval $[-1,1]$, is 
presented through a result established by \cite{King:2001} (refer to the Appendix).

The following result establishes the invariance of the proposed coefficient of concordance for elliptically contoured 
distributions when $\mu_1=\mu_2$ and the discrepancy between $X_1$ and $X_2$ is measured in an $L_p$ space.
\begin{proposition}\label{prop:2}
  Assume that $(X_1, X_2)^\top \sim \mathsf{EC}_2(\bs{\mu}, \bs{\Sigma};g)$ with $\mu_1 = \mu_2$, and suppose that 
  $\varphi(z)=\lvert z\rvert^p, p\in\mathbb{N}$. Then
  \begin{equation}\label{eq:Lp}
    \rho_p = 1 - \frac{\E(\lvert X_1 - X_2\rvert^p)}{\E(\lvert X_1 - X_2\rvert^p: \sigma_{12} = 0)} = 1 - (1 - \rho_c)^{p/2}.
  \end{equation}
\end{proposition}

Equation \eqref{eq:Lp} establishes a relationship between $\rho_p$ and $\rho_c$ that significantly simplifies the 
estimation process of these concordance coefficients when employing a plug-in type estimation method.

In the next section we discussed a criterion to select $p$ in an optimal way by minimizing the asymptotic variance 
of a sample version of $\rho_p$.

\subsection{Inference}\label{sec:inference}

In order to establish statistical properties of the sample counterpart of coefficient $\rho_1$ that lead us to 
approximate confidence intervals or hypothesis testing problems, when the underlying distribution of the sample 
points is normal, that is $(X_{11},X_{21})^\top,\dots,(X_{1n},X_{2n})^\top$ is a random sample from $\mathsf{N}_2(\bs{\mu},
\bs{\Sigma})$ and assuming that $\mu_1 = \mu_2$. In this case, by plugging in, a sample version of $\rho_p$ is
\[
  \what{\rho}_p = 1 - (1 - \what{\rho}_c)^{p/2}.
\]
Again, by the delta method, but using $z\mapsto 1-\sqrt{1-z}$ we obtain the asymptotic distribution of $\what{\rho}_1$, 
\begin{equation}\label{eq:asymp-rho1}
  \sqrt{n}(\what{\rho}_1 - \rho_1) \stackrel{D}{\rightarrow} \mathsf{N}\big(0, v^2(1-\rho_c^2)^2(\half(1-\rho_c)^{-1/2})^2\big),
\end{equation}
as $n\to\infty$. Moreover, this can be easily generalized to $\what{\rho}_p$ when $\mu_1 = \mu_2$ through the map 
$z\mapsto 1 - (1 - z)^{p/2}$. Then,
\[
  \sqrt{n}(\what{\rho}_p - \rho_p) \stackrel{D}{\rightarrow} \mathsf{N}\big(0, v^2(1-\rho_c^2)^2(\textstyle\frac{p}{2}
  (1-\rho_c)^{p/2-1})^2\big),
\]
as $n\to\infty$.

\begin{proposition}\label{prop:3} 
  Assume that $(X_{11},X_{21})^\top,\dots,(X_{1n},X_{2n})^\top$ is a random sample from $\mathsf{EC}_2(\bs{\mu},\bs{\Sigma};\psi)$,
  with $\mu_1 = \mu_2$. Then the asymptotic variance, $\var(\what{\rho}_p)$, attains a minimum at $p=1$, for $\rho < 0$.
\end{proposition}

The provided information clarifies the choice between Lin's coefficient ($\rho_2$) and the $L_1$ coefficient ($\rho_1$) 
when dealing with a bivariate EC distribution. It is evident that this choice is contingent upon the correlation value 
between variables $X$ and $Y$.

\begin{proposition}\label{prop:4} 
  Assume that $(X_{11},X_{21})^\top,\dots,(X_{1n},X_{2n})^\top$ is a random sample from $\mathsf{EC}_2(\bs{\mu},\bs{\Sigma};\psi)$,
  with $\mu_1=\mu_2$. Then $\var(\what{\rho}_1) \leq \var(\what{\rho}_2)$ if and only if $\rho\leq \min\big\{1, \textstyle
  \frac{3}{8}(\sigma_{11} + \sigma_{22})/\sqrt{\sigma_{11}\sigma_{22}}\big\}$.
\end{proposition}

\begin{proof} 
  If $\var(\what{\rho}_1) \leq \var(\what{\rho}_2)$, it follows that 
  \[
    \frac{1}{4} \leq 1 - \frac{2\rho\sqrt{\sigma_{11}\sigma_{22}}}{\sigma_{11} + \sigma_{22}},
  \]
  and therefore $\rho\leq\textstyle\frac{3}{8}(\sigma_{11} + \sigma_{22})/\sqrt{\sigma_{11}\sigma_{22}}$. 
  The converse is analogous.
\end{proof}

\begin{figure}[!ht]
  \vskip -1.75em
  \centering
  \includegraphics[width = 0.45\linewidth]{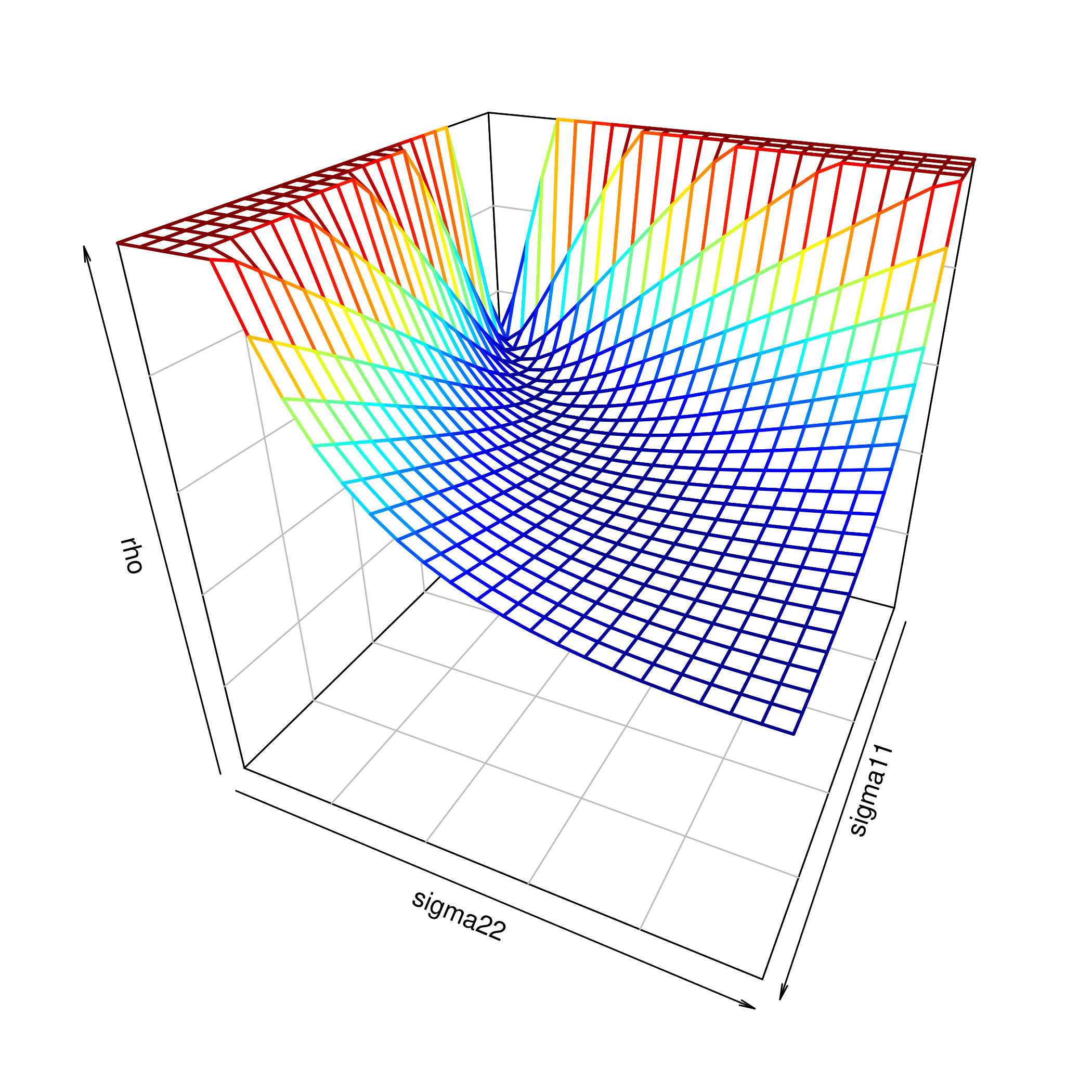}
  \caption{Plot of upper bound for $\rho$ given in Proposition \ref{prop:4}.}
\end{figure}

To offer some degree of protection against outliers, we will focus on modeling in the context of multivariate 
analysis using the Laplace distribution, a member from the class of elliptically contoured distributions that has 
been studied in detail, for example in \cite{Gomez:1998, Kotz:2001, Gomez:2008}. More recently, the Laplace distribution 
has been proposed for robust estimation in the linear mixed-effects model by \cite{Yavuz:2018}, who developed an 
EM algorithm to perform the parameter estimation. Following \cite{Yavuz:2018} (see also \cite{Pinheiro:2001}) we 
obtain an efficient procedure for parameter estimation based on a random sample $\bs{x}_1,\dots,\bs{x}_n$ from the 
$\mathsf{Laplace}_k(\bs{\mu},\bs{\Sigma})$ distribution, with density
\[
  f(\bs{x}_i;\bs{\mu},\bs{\Sigma}) = \frac{\Gamma(k/2)}{\pi^{k/2}\Gamma(k)2^{k+1}}\,\lvert\bs{\Sigma}\rvert^{-1/2}
  \exp\big[-\half\big\{(\bs{x}_i - \bs{\mu})^\top\bs{\Sigma}^{-1}(\bs{x}_i - \bs{\mu})\big\}^{1/2}\big], \qquad 
  i = 1,\dots,n,
\]
by applying a data augmentation scheme incorporating latent variables $\omega_1,\dots,\omega_n$, and using an 
EM algorithm\cite{Dempster:1977} based on the hierarchical representation $\bs{X}_i\mid\omega_i \sim \mathsf{N}_k(\bs{\mu},
\bs{\Sigma})$, with $\omega_i$ being a positive random variable having a probability density function
\begin{equation}\label{eq:mixture}
  h(\omega_i) = \frac{1}{2^{(3k+1)/2}\Gamma\big(\frac{k+1}{2}\big)}\,\omega_i^k \exp(-\omega_i^2/8), 
  \quad \omega_i > 0,
\end{equation}
for $i=1,\dots,n$. It is easy to notice that, disregarding terms that do not depend on $\bs{\theta} = (\bs{\mu}^\top,
\bs{\sigma}^\top)^\top$, with $\bs{\sigma} = \vech(\bs{\Sigma})$ denoting the different elements in the $\bs{\Sigma}$ 
matrix, the conditional expectation of the complete-data log-likelihood function, 
takes the form.
\[
  Q(\bs{\theta};\bs{\theta}^{(r)}) = -\frac{n}{2}\log\lvert\bs{\Sigma}\rvert - \frac{1}{2}\sum_{i=1}^n \omega_i^{(r)}
  (\bs{x}_i - \bs{\mu})^\top\bs{\Sigma}^{-1}(\bs{x}_i - \bs{\mu}),
\]
where
\begin{equation}\label{eq:E-step}
  \omega_i^{(r)} = \E(\omega_i^{-2}\mid\bs{x}_i,\bs{\theta}^{(r)}) = \frac{1/2}{D_i(\bs{\theta}^{(r)})}\,
  \frac{K_{k/2-1}(D_i(\bs{\theta}^{(r)})/2)}{K_{k/2}(D_i(\bs{\theta}^{(r)})/2)}, \qquad i = 1,\dots,n,
\end{equation}
with $D_i^2(\bs{\theta}^{(r)}) = (\bs{x}_i - \bs{\mu}^{(r)})^\top\{\bs{\Sigma}^{(r)}\}^{-1}(\bs{x}_i - \bs{\mu}^{(r)})$,
while $K_\nu(\cdot)$ represents the modified Bessel function of second type of order $\nu$ and $\bs{\theta}^{(r)}$ 
denotes an initial estimate for the $r$-th stage of the algorithm. Maximizing $Q(\bs{\theta};\bs{\theta}^{(r)})$ with 
respect to $\bs{\mu}$ and $\bs{\sigma}$ we obtain the following updates,
\begin{align} 
  \bs{\mu}^{(r+1)} & = \frac{1}{\sum_{j=1}^n \omega_j^{(r)}}\,\sum_{i=1}^n \omega_i^{(r)}\bs{x}_i, \label{eq:M-center} \\
  \bs{\Sigma}^{(r+1)} & = \frac{1}{n}\sum_{i=1}^n \omega_i^{(r)}(\bs{x}_i - \bs{\mu}^{(r+1)})
  (\bs{x}_i - \bs{\mu}^{(r+1)})^\top. \label{eq:M-scatter}
\end{align}
The steps described in Equations \eqref{eq:E-step}, \eqref{eq:M-center} and \eqref{eq:M-scatter}, must be iterated 
until convergence is reached. We must emphasize that in our numerical experiments we will have $k = 2$ measuring 
instruments. On the other hand, to deal with the parameter estimation subject to the constraint $\mu_1 = \mu_2$, 
a rather simple modification of the previous procedure must be considered, the details of which are presented in 
the Appendix.

Let $\what{\bs{\mu}}$ and $\what{\bs{\Sigma}}$ be the ML estimators for a random sample from a Laplace distribution 
obtained, for example, from the procedure described in Equations \eqref{eq:E-step}-\eqref{eq:M-scatter}. In the general 
context of elliptical contoured distributions, the Fisher information matrix required to characterize the asymptotic 
distribution of the ML estimators has been obtained independently in \cite{Tyler:1982} and \cite{Mitchell:1989}. Thus, 
it follows that
\[
  \sqrt{n}(\what{\bs{\mu}} - \bs{\mu}) \stackrel{D}{\longrightarrow} \mathsf{N}_k\Big(\bs{0},4k\bs{\Sigma}\Big).
\]

Because of the important simplifications that occur in the case that $\mu_1 = \mu_2$. Next, we will be interested 
in testing linear hypotheses of the form $H_0:\bs{A\mu} = \bs{0}$, where $\bs{A}$ is a known matrix of order $r \times k$, 
with $rk(\bs{A}) = r$ $(r \leq k)$. To test this type of hypothesis we shall consider tests for large samples based 
on the likelihood, namely the Wald\cite{Wald:1943}, Rao score\cite{Rao:1948} and gradient statistics\cite{Terrell:2002}. 
Those that take the form:
\begin{align} 
  W_n & = \frac{n}{4k}\what{\bs{\mu}}{}^\top\bs{A}^\top(\bs{A}\what{\bs{\Sigma}}\bs{A}^\top)^{-1}\bs{A}\what{\bs{\mu}}, \label{eq:Wald} \\
  R_n & = \frac{4k}{n}\Big\{\sum_{i=1}^n\omega_i(\widetilde{\bs{\theta}})(\bs{x}_i - \widetilde{\bs{\mu}})\Big\}^\top
  \widetilde{\bs{\Sigma}}{}^{-1}\Big\{\sum_{i=1}^n\omega_i(\widetilde{\bs{\theta}})(\bs{x}_i - \widetilde{\bs{\mu}})\Big\}, \label{eq:Rao} \\
  G_n & = \Big\{\sum_{i=1}^n\omega_i(\widetilde{\bs{\theta}})(\bs{x}_i - \widetilde{\bs{\mu}})\Big\}^\top\widetilde{\bs{\Sigma}}{}^{-1}
  \what{\bs{\mu}}, \label{eq:Terrell}
\end{align}
where $\widetilde{\bs{\theta}}$ and $\what{\bs{\theta}}$ are the restricted and unrestricted ML estimators, respectively.
We should note that $\omega_i(\widetilde{\bs{\theta}})$ correspond to the estimated weights obtained at the convergence 
of the algorithm for constrained estimation outlined in the Appendix. Indeed, under $H_0:\bs{A\mu} = \bs{0}$, all of these 
test statistics are asymptotically equivalent, each with chi-squared distribution with $r$ degrees of freedom. Furthermore, 
we can appreciate that, both $\widetilde{\bs{\Sigma}}$ and $\what{\bs{\Sigma}}$ are consistent estimators of $\bs{\Sigma}$.

An interesting alternative for testing hypotheses of the type $H_0:\bs{A\mu} = \bs{0}$ which is particularly useful as it 
does not require iterative algorithms, is based on $\bs{x}_1,\dots,\bs{x}_n$ a random sample from a population with a 
$\mathsf{Laplace}_k(\bs{\mu},\bs{\Sigma})$ distribution, and consider $\overline{\bs{x}}=\frac{1}{n}\sum_{i=1}^n\bs{x}_i$ 
and $\bs{S}=\frac{1}{n-1}\sum_{i=1}^n(\bs{x}_i-\overline{\bs{x}})(\bs{x}_i-\overline{\bs{x}})^\top$ as the mean vector 
and covariance matrix of sample size $n$, respectively. We know that $\overline{\bs{x}}$ and $\bs{S}$ are unbiased and 
consistent estimators of $\bs{\mu}$ and $\bs{\Sigma}$, respectively. In addition, $\overline{\bs{x}}$ and $\bs{S}$ are 
asymptotically independent, with limit distributions given, respectively, by
\[
  \sqrt{n}(\overline{\bs{x}}-\bs{\mu}) \stackrel{D}{\longrightarrow} \mathsf{N}_k(\bs{0}, \bs{\Sigma}), \qquad
  \sqrt{n}(\vech(\bs{S})-\vech(\bs{\Sigma})) \stackrel{D}{\longrightarrow} \mathsf{N}_{k^*}(\bs{0}, \bs{\Omega}),
\]
where $k^*=k(k+1)/2$ and
\[
 \bs{\Omega} = \frac{1}{k+1}\bs{D}_k^\top\{(k+3)(\bs{I}_{k^2}+\bs{K}_k)(\bs{\Sigma}\otimes\bs{\Sigma}) + 2(\ovec\bs{\Sigma})
 (\ovec\bs{\Sigma})^\top\}\bs{D}_k,
\]
with $\bs{D}_k$ and $\bs{K}_k$ being duplication and commutation matrices of order $k$, respectively \cite{Anderson:1992, 
Anderson:2003}. It is straightforward to note that the excess kurtosis associated with a $\mathsf{Laplace}_k(\bs{\mu},\bs{\Sigma})$ 
distribution is given by $\kappa=2/(k+1)$.The above result is valid for the whole class of multivariate elliptical distributions 
and enables the use of the generalized Hotelling's $T^2$ statistic, which adopts the form
\begin{equation}\label{eq:T2}
  T_n^2 = n\overline{\bs{x}}{}^\top\bs{A}^\top(\bs{ASA}^\top)^{-1}\bs{A}\overline{\bs{x}}.
\end{equation}
Although its small sample distribution is difficult to obtain, the asymptotic distribution of $T_n^2$ can be obtained 
by noticing that $\sqrt{n}\bs{A}(\overline{\bs{x}}-\bs{\mu})\stackrel{D}{\longrightarrow}\mathsf{N}_r(\bs{0},\bs{A\Sigma}
\bs{A}^\top)$ and $\bs{S} \stackrel{p}{\rightarrow} \bs{\Sigma}$ \cite{Anderson:2003}. This leads to notice that, under 
$H_0$, the test statistic in \eqref{eq:T2} follows an asymptotically chi-squared distribution with $r$ degrees of freedom. 
We can emphasize that the main advantage of the generalized Hotelling $T^2$ statistic is that it does not require any 
iterative algorithm to obtain parameter estimates, instead it is based on moment estimators.

\subsection{Goodness of fit}\label{sec:gof}

Following \cite{Osorio:2023}, we can use the Wilson-Hilferty\cite{Wilson:1931,Terrell:2003} transformation to construct 
a graphical device to evaluate goodness-of-fit. This model checking mechanism is based on noting that for $\bs{x}_i\sim
\mathsf{Laplace}_k(\bs{\mu},\bs{\Sigma})$, we have that the Mahalanobis distances satisfy $D_i(\bs{\theta})\sim\mathsf{Gamma}(k,1/2)$ 
for $i=1,\dots,n$. This result follows from \cite{Gomez:1998} and allows the construction of QQ-plots with envelopes by 
realizing that the variables
\[
  z_i = \frac{\what{F}_i^{1/3} - \big(1 - \frac{1}{9k}\big)}{1/\sqrt{9k}},
  \qquad i=1,\dots,n,
\]
approximately follow a standard normal distribution, where $\what{F}_i = D_i(\what{\bs{\theta}})/(2k)$. This result also 
allows us to use the transformed variables $\{z_1,\dots,z_n\}$ to evaluate the distributional assumption by applying a 
omnibus tests for normality proposed, for instance, in \cite{Shapiro:1965} or \cite{Jarque:1980}. In addition, the modified 
distances $\what{F}_i$, for $i=1,\dots,n$ enable the identification of outlying observations.

\begin{table*}[!ht]%
  \caption{Averages of Lin's, $L_1$ and $U$-statistics based concordance correlation coefficients estimates under Scenarios a), b) and c).\label{tab:est01}}
  \begin{tabular*}{\textwidth}{@{\extracolsep\fill}cclcccccccccc@{}} \toprule
  Scenario & $m$ & Fitted & \multicolumn{3}{@{}c}{$\what{\rho}_c$} & \multicolumn{3}{@{}c}{$\what{\rho}_1$} & $\varphi$ & \multicolumn{3}{@{}c}{$\what{\rho}_\varphi$} \\ \cmidrule{4-6}\cmidrule{7-9}\cmidrule{11-13}
     &   & Model    & 25    & 100   & 400   & 25    & 100   & 400   &             & 25    & 100   & 400   \\ \midrule
  a) & 1 & Gaussian & 0.945 & 0.948 & 0.950 & 0.769 & 0.773 & 0.776 & $(\cdot)^2$ & 0.945 & 0.948 & 0.950 \\
     &   & Laplace  & 0.944 & 0.948 & 0.950 & 0.768 & 0.773 & 0.776 & $|\cdot|$   & 0.766 & 0.773 & 0.776 \\
     & 2 & Gaussian & 0.834 & 0.848 & 0.850 & 0.600 & 0.612 & 0.613 & $(\cdot)^2$ & 0.834 & 0.848 & 0.850 \\
     &   & Laplace  & 0.832 & 0.848 & 0.850 & 0.598 & 0.611 & 0.613 & $|\cdot|$   & 0.598 & 0.611 & 0.613 \\
     & 3 & Gaussian & 0.722 & 0.746 & 0.748 & 0.481 & 0.498 & 0.499 & $(\cdot)^2$ & 0.724 & 0.747 & 0.749 \\
     &   & Laplace  & 0.720 & 0.746 & 0.748 & 0.479 & 0.498 & 0.499 & $|\cdot|$   & 0.478 & 0.498 & 0.499 \\ \midrule
  b) & 1 & Gaussian & 0.940 & 0.948 & 0.950 & 0.763 & 0.773 & 0.776 & $(\cdot)^2$ & 0.940 & 0.948 & 0.950 \\
     &   & Laplace  & 0.941 & 0.948 & 0.950 & 0.763 & 0.773 & 0.776 & $|\cdot|$   & 0.770 & 0.782 & 0.785 \\
     & 2 & Gaussian & 0.828 & 0.845 & 0.849 & 0.595 & 0.609 & 0.612 & $(\cdot)^2$ & 0.828 & 0.845 & 0.849 \\
     &   & Laplace  & 0.830 & 0.846 & 0.849 & 0.597 & 0.609 & 0.612 & $|\cdot|$   & 0.609 & 0.624 & 0.627 \\
     & 3 & Gaussian & 0.718 & 0.744 & 0.749 & 0.481 & 0.497 & 0.499 & $(\cdot)^2$ & 0.719 & 0.744 & 0.749 \\
     &   & Laplace  & 0.720 & 0.745 & 0.749 & 0.482 & 0.498 & 0.499 & $|\cdot|$   & 0.497 & 0.517 & 0.519 \\ \midrule
  c) & 1 & Gaussian & 0.877 & 0.885 & 0.882 & 0.718 & 0.724 & 0.723 & $(\cdot)^2$ & 0.877 & 0.885 & 0.882 \\
     &   & Laplace  & 0.903 & 0.919 & 0.922 & 0.734 & 0.750 & 0.753 & $|\cdot|$   & 0.781 & 0.803 & 0.810 \\
     & 2 & Gaussian & 0.742 & 0.742 & 0.741 & 0.558 & 0.562 & 0.561 & $(\cdot)^2$ & 0.742 & 0.746 & 0.741 \\
     &   & Laplace  & 0.777 & 0.777 & 0.802 & 0.575 & 0.585 & 0.588 & $|\cdot|$   & 0.638 & 0.667 & 0.680 \\
     & 3 & Gaussian & 0.605 & 0.605 & 0.621 & 0.451 & 0.447 & 0.452 & $(\cdot)^2$ & 0.608 & 0.611 & 0.621 \\
     &   & Laplace  & 0.650 & 0.652 & 0.702 & 0.466 & 0.468 & 0.482 & $|\cdot|$   & 0.551 & 0.573 & 0.595 \\
  \bottomrule
  \end{tabular*}
\end{table*}

\begin{table*}[!ht]%
  \caption{Averages of the standard errors of Lin's, $L_1$ and $U$-statistics based concordance correlation coefficients estimates under Scenarios a), b) and c).\label{tab:est02}}
  \begin{tabular*}{\textwidth}{@{\extracolsep\fill}cclcccccccccc@{}} \toprule
  Scenario & $m$ & Fitted & \multicolumn{3}{@{}c}{$\SE(\what{\rho}_c)$} & \multicolumn{3}{@{}c}{$\SE(\what{\rho}_1)$} & $\varphi$ & \multicolumn{3}{@{}c}{$\SE(\what{\rho}_\varphi)$} \\ \cmidrule{4-6}\cmidrule{7-9}\cmidrule{11-13}
     &   & Model    & 25    & 100   & 400   & 25    & 100   & 400   &             & 25    & 100   & 400   \\ \midrule
  a) & 1 & Gaussian & 0.022 & 0.010 & 0.005 & 0.046 & 0.022 & 0.011 & $(\cdot)^2$ & 0.019 & 0.010 & 0.005 \\
     &   & Laplace  & 0.011 & 0.005 & 0.003 & 0.046 & 0.022 & 0.011 & $|\cdot|$   & 0.044 & 0.023 & 0.011 \\
     & 2 & Gaussian & 0.061 & 0.028 & 0.014 & 0.074 & 0.036 & 0.018 & $(\cdot)^2$ & 0.074 & 0.036 & 0.018 \\
     &   & Laplace  & 0.075 & 0.036 & 0.018 & 0.075 & 0.036 & 0.018 & $|\cdot|$   & 0.072 & 0.037 & 0.019 \\
     & 3 & Gaussian & 0.096 & 0.044 & 0.023 & 0.090 & 0.044 & 0.022 & $(\cdot)^2$ & 0.085 & 0.043 & 0.022 \\
     &   & Laplace  & 0.090 & 0.044 & 0.022 & 0.090 & 0.044 & 0.022 & $|\cdot|$   & 0.088 & 0.045 & 0.023 \\ \midrule
  b) & 1 & Gaussian & 0.023 & 0.010 & 0.005 & 0.046 & 0.022 & 0.011 & $(\cdot)^2$ & 0.024 & 0.012 & 0.006 \\
     &   & Laplace  & 0.012 & 0.005 & 0.003 & 0.046 & 0.022 & 0.011 & $|\cdot|$   & 0.048 & 0.024 & 0.012 \\
     & 2 & Gaussian & 0.062 & 0.029 & 0.014 & 0.074 & 0.036 & 0.018 & $(\cdot)^2$ & 0.063 & 0.036 & 0.018 \\
     &   & Laplace  & 0.031 & 0.014 & 0.007 & 0.074 & 0.036 & 0.018 & $|\cdot|$   & 0.075 & 0.040 & 0.020 \\
     & 3 & Gaussian & 0.095 & 0.044 & 0.022 & 0.088 & 0.044 & 0.022 & $(\cdot)^2$ & 0.094 & 0.053 & 0.028 \\
     &   & Laplace  & 0.048 & 0.022 & 0.011 & 0.089 & 0.044 & 0.022 & $|\cdot|$   & 0.090 & 0.048 & 0.024 \\ \midrule
  c) & 1 & Gaussian & 0.026 & 0.013 & 0.006 & 0.037 & 0.018 & 0.009 & $(\cdot)^2$ & 0.041 & 0.040 & 0.040 \\
     &   & Laplace  & 0.013 & 0.006 & 0.003 & 0.041 & 0.021 & 0.010 & $|\cdot|$   & 0.057 & 0.045 & 0.037 \\
     & 2 & Gaussian & 0.055 & 0.027 & 0.013 & 0.054 & 0.027 & 0.013 & $(\cdot)^2$ & 0.079 & 0.082 & 0.079 \\
     &   & Laplace  & 0.029 & 0.014 & 0.007 & 0.062 & 0.032 & 0.016 & $|\cdot|$   & 0.084 & 0.068 & 0.058 \\
     & 3 & Gaussian & 0.070 & 0.036 & 0.018 & 0.059 & 0.030 & 0.015 & $(\cdot)^2$ & 0.101 & 0.111 & 0.112 \\
     &   & Laplace  & 0.041 & 0.021 & 0.011 & 0.071 & 0.037 & 0.020 & $|\cdot|$   & 0.096 & 0.082 & 0.069 \\
  \bottomrule
  \end{tabular*}
\end{table*}

\section{Numerical experiments}

In this section we illustrate the proposed methodology through the analysis of a real-life dataset. In addition, we present 
a simulation study to evaluate the performance of the proposed concordance coefficient. {\R} codes for the estimation of Lin's 
and $L_1$ coefficients, large samples test statistics and the Wilson-Hilferty transformation for goodness-of-fit described in 
the previous sections are available on {\github}.\footnote{URL:\,\url{https://github.com/faosorios/L1CCC}}. In our analyses we 
also have used routines available in the \textsf{L1pack}\cite{L1pack} and \textsf{fastmatrix}\citep{fastmatrix} packages. 

\begin{table*}[!ht]%
  \caption{Averages of Lin's, $L_1$ and $U$-statistics based concordance correlation coefficients estimates under Scenario d).\label{tab:est03}}
  \begin{tabular*}{\textwidth}{@{\extracolsep\fill}ccclcccccccccc@{}} \toprule
  $m$ & $\eta$ & $\epsilon$ & Fitted & \multicolumn{3}{@{}c}{$\what{\rho}_c$} & \multicolumn{3}{@{}c}{$\what{\rho}_1$} & $\varphi$ & \multicolumn{3}{@{}c}{$\what{\rho}_\varphi$} \\ \cmidrule{5-7}\cmidrule{8-10}\cmidrule{12-14}
      &    &      & Model    & 25    & 100   & 400   & 25    & 100   & 400   &             & 25    & 100   & 400   \\ \midrule
  1   &  5 & 0.05 & Gaussian & 0.934 & 0.944 & 0.949 & 0.759 & 0.770 & 0.775 & $(\cdot)^2$ & 0.933 & 0.944 & 0.949 \\
      &    &      & Laplace  & 0.940 & 0.948 & 0.950 & 0.764 & 0.774 & 0.776 & $|\cdot|$   & 0.770 & 0.783 & 0.786 \\
      &    & 0.10 & Gaussian & 0.930 & 0.943 & 0.949 & 0.754 & 0.768 & 0.775 & $(\cdot)^2$ & 0.929 & 0.943 & 0.949 \\
      &    &      & Laplace  & 0.939 & 0.947 & 0.949 & 0.762 & 0.772 & 0.776 & $|\cdot|$   & 0.774 & 0.787 & 0.791 \\
      &    & 0.25 & Gaussian & 0.934 & 0.945 & 0.949 & 0.758 & 0.770 & 0.775 & $(\cdot)^2$ & 0.933 & 0.945 & 0.949 \\
      &    &      & Laplace  & 0.939 & 0.947 & 0.949 & 0.764 & 0.772 & 0.776 & $|\cdot|$   & 0.783 & 0.794 & 0.798 \\ \midrule
      & 10 & 0.05 & Gaussian & 0.915 & 0.936 & 0.947 & 0.745 & 0.761 & 0.773 & $(\cdot)^2$ & 0.915 & 0.936 & 0.947 \\
      &    &      & Laplace  & 0.934 & 0.947 & 0.949 & 0.759 & 0.772 & 0.776 & $|\cdot|$   & 0.775 & 0.792 & 0.797 \\
      &    & 0.10 & Gaussian & 0.909 & 0.939 & 0.948 & 0.737 & 0.763 & 0.774 & $(\cdot)^2$ & 0.909 & 0.939 & 0.948 \\
      &    &      & Laplace  & 0.931 & 0.945 & 0.949 & 0.755 & 0.770 & 0.775 & $|\cdot|$   & 0.780 & 0.798 & 0.804 \\
      &    & 0.25 & Gaussian & 0.928 & 0.944 & 0.949 & 0.754 & 0.769 & 0.775 & $(\cdot)^2$ & 0.928 & 0.944 & 0.949 \\
      &    &      & Laplace  & 0.935 & 0.946 & 0.949 & 0.761 & 0.771 & 0.776 & $|\cdot|$   & 0.793 & 0.805 & 0.810 \\ \midrule
  2   &  5 & 0.05 & Gaussian & 0.813 & 0.835 & 0.847 & 0.590 & 0.603 & 0.612 & $(\cdot)^2$ & 0.813 & 0.835 & 0.847 \\
      &    &      & Laplace  & 0.829 & 0.844 & 0.850 & 0.598 & 0.608 & 0.613 & $|\cdot|$   & 0.610 & 0.623 & 0.630 \\
      &    & 0.10 & Gaussian & 0.808 & 0.834 & 0.847 & 0.587 & 0.600 & 0.612 & $(\cdot)^2$ & 0.808 & 0.834 & 0.847 \\
      &    &      & Laplace  & 0.826 & 0.842 & 0.849 & 0.596 & 0.606 & 0.613 & $|\cdot|$   & 0.618 & 0.632 & 0.640 \\
      &    & 0.25 & Gaussian & 0.820 & 0.839 & 0.848 & 0.596 & 0.604 & 0.611 & $(\cdot)^2$ & 0.820 & 0.839 & 0.848 \\
      &    &      & Laplace  & 0.830 & 0.843 & 0.849 & 0.601 & 0.607 & 0.612 & $|\cdot|$   & 0.635 & 0.645 & 0.651 \\ \midrule
      & 10 & 0.05 & Gaussian & 0.771 & 0.817 & 0.843 & 0.570 & 0.593 & 0.610 & $(\cdot)^2$ & 0.771 & 0.817 & 0.843 \\
      &    &      & Laplace  & 0.815 & 0.840 & 0.849 & 0.591 & 0.606 & 0.613 & $|\cdot|$   & 0.617 & 0.639 & 0.649 \\
      &    & 0.10 & Gaussian & 0.772 & 0.825 & 0.846 & 0.569 & 0.595 & 0.611 & $(\cdot)^2$ & 0.772 & 0.825 & 0.846 \\
      &    &      & Laplace  & 0.811 & 0.838 & 0.849 & 0.588 & 0.603 & 0.613 & $|\cdot|$   & 0.631 & 0.652 & 0.663 \\
      &    & 0.25 & Gaussian & 0.811 & 0.837 & 0.847 & 0.592 & 0.603 & 0.611 & $(\cdot)^2$ & 0.811 & 0.837 & 0.847 \\
      &    &      & Laplace  & 0.824 & 0.841 & 0.848 & 0.599 & 0.606 & 0.612 & $|\cdot|$   & 0.653 & 0.665 & 0.672 \\ \midrule
  3   &  5 & 0.05 & Gaussian & 0.701 & 0.729 & 0.743 & 0.476 & 0.489 & 0.496 & $(\cdot)^2$ & 0.702 & 0.730 & 0.743 \\
      &    &      & Laplace  & 0.715 & 0.741 & 0.748 & 0.479 & 0.494 & 0.498 & $|\cdot|$   & 0.496 & 0.515 & 0.521 \\
      &    & 0.10 & Gaussian & 0.690 & 0.729 & 0.746 & 0.471 & 0.488 & 0.498 & $(\cdot)^2$ & 0.691 & 0.729 & 0.746 \\
      &    &      & Laplace  & 0.711 & 0.739 & 0.748 & 0.477 & 0.493 & 0.499 & $|\cdot|$   & 0.507 & 0.527 & 0.535 \\
      &    & 0.25 & Gaussian & 0.701 & 0.738 & 0.746 & 0.478 & 0.493 & 0.498 & $(\cdot)^2$ & 0.703 & 0.738 & 0.746 \\
      &    &      & Laplace  & 0.713 & 0.741 & 0.747 & 0.481 & 0.494 & 0.498 & $|\cdot|$   & 0.525 & 0.544 & 0.549 \\ \midrule
      & 10 & 0.05 & Gaussian & 0.665 & 0.705 & 0.736 & 0.465 & 0.480 & 0.493 & $(\cdot)^2$ & 0.667 & 0.705 & 0.737 \\
      &    &      & Laplace  & 0.702 & 0.736 & 0.746 & 0.476 & 0.492 & 0.498 & $|\cdot|$   & 0.509 & 0.536 & 0.545 \\
      &    & 0.10 & Gaussian & 0.652 & 0.716 & 0.743 & 0.457 & 0.483 & 0.497 & $(\cdot)^2$ & 0.653 & 0.716 & 0.743 \\
      &    &      & Laplace  & 0.695 & 0.733 & 0.747 & 0.473 & 0.490 & 0.498 & $|\cdot|$   & 0.526 & 0.553 & 0.564 \\
      &    & 0.25 & Gaussian & 0.689 & 0.735 & 0.745 & 0.474 & 0.492 & 0.497 & $(\cdot)^2$ & 0.691 & 0.736 & 0.745 \\
      &    &      & Laplace  & 0.705 & 0.738 & 0.747 & 0.480 & 0.494 & 0.498 & $|\cdot|$   & 0.550 & 0.570 & 0.576 \\
  \bottomrule
  \end{tabular*}
\end{table*}

\subsection{Monte Carlo simulation study}

We conducted a Monte Carlo simulation study designed to investigate the performance of the coefficients $\what{\rho}_c$ 
and $\what{\rho}_1$, as well as to examine the empirical size of the hypothesis test statistics to assess the equality 
of means $H_0: \mu_1 = \mu_2$. In our experiment, 1000 datasets with sample size $n = 25,100$, and $400$ were created from 
the following scenarios:
\[
  \mathrm{a)}\ \mathsf{N}_2(\bs{0},\bs{\Sigma}_m), \qquad 
  \mathrm{b)}\ \mathsf{Laplace}_2(\bs{0},\bs{\Sigma}_m), \qquad 
  \mathrm{c)}\ \mathsf{Cauchy}_2(\bs{0},\bs{\Sigma}_m), \qquad 
  \mathrm{d)}\ \mathsf{CN}_2(\bs{0},\bs{\Sigma}_m,\epsilon,\eta),
\]
for $m=1,2,3$, with
\[
  \bs{\Sigma}_1 = \begin{pmatrix}
    1.00 & 0.95 \\
    0.95 & 1.00
  \end{pmatrix}, \qquad \bs{\Sigma}_2 = \begin{pmatrix}
    1.00 & 0.85 \\
    0.85 & 1.00
  \end{pmatrix}, \qquad \bs{\Sigma}_3 = \begin{pmatrix}
    1.00 & 0.75 \\
    0.75 & 1.00
  \end{pmatrix},
\]
where $\epsilon = 0.05,0.10,0.25$ denotes the percentage of contamination and $\eta = 5,10$ represents the variance 
inflation factor of a bivariate contaminated normal distribution.  Our simulation study has been 
carried out under the assumption that $\mu_1 = \mu_2$, leading to the following reference values, $\rho_c = 0.95, 
0.85, 0.75$ and $\rho_1 = 0.78, 0.61, 0.50$ for cases $m = 1,2,3$, respectively.

\begin{table*}[!ht]%
  \caption{Averages of the standard errors of Lin's, $L_1$ and $U$-statistics based concordance correlation coefficients estimates under Scenario d).\label{tab:est04}}
  \begin{tabular*}{\textwidth}{@{\extracolsep\fill}ccclcccccccccc@{}} \toprule
  $m$ & $\eta$ & $\epsilon$ & Fitted & \multicolumn{3}{@{}c}{$\SE(\what{\rho}_c)$} & \multicolumn{3}{@{}c}{$\SE(\what{\rho}_1)$} & $\varphi$ & \multicolumn{3}{@{}c}{$\SE(\what{\rho}_\varphi)$} \\ \cmidrule{5-7}\cmidrule{8-10}\cmidrule{12-14}
      &    &      & Model    & 25    & 100   & 400   & 25    & 100   & 400   &             & 25    & 100   & 400   \\ \midrule
  1   &  5 & 0.05 & Gaussian & 0.024 & 0.011 & 0.005 & 0.045 & 0.022 & 0.011 & $(\cdot)^2$ & 0.028 & 0.019 & 0.012 \\
      &    &      & Laplace  & 0.012 & 0.005 & 0.003 & 0.046 & 0.022 & 0.011 & $|\cdot|$   & 0.049 & 0.027 & 0.014 \\
      &    & 0.10 & Gaussian & 0.025 & 0.011 & 0.005 & 0.045 & 0.022 & 0.011 & $(\cdot)^2$ & 0.033 & 0.021 & 0.011 \\
      &    &      & Laplace  & 0.012 & 0.005 & 0.003 & 0.046 & 0.022 & 0.011 & $|\cdot|$   & 0.052 & 0.029 & 0.015 \\
      &    & 0.25 & Gaussian & 0.024 & 0.011 & 0.005 & 0.045 & 0.022 & 0.011 & $(\cdot)^2$ & 0.032 & 0.017 & 0.009 \\
      &    &      & Laplace  & 0.012 & 0.005 & 0.003 & 0.046 & 0.022 & 0.011 & $|\cdot|$   & 0.053 & 0.029 & 0.014 \\ \midrule
      & 10 & 0.05 & Gaussian & 0.026 & 0.011 & 0.005 & 0.042 & 0.022 & 0.011 & $(\cdot)^2$ & 0.036 & 0.030 & 0.018 \\
      &    &      & Laplace  & 0.012 & 0.005 & 0.003 & 0.045 & 0.022 & 0.011 & $|\cdot|$   & 0.053 & 0.034 & 0.019 \\
      &    & 0.10 & Gaussian & 0.029 & 0.011 & 0.005 & 0.043 & 0.022 & 0.011 & $(\cdot)^2$ & 0.044 & 0.028 & 0.014 \\
      &    &      & Laplace  & 0.013 & 0.005 & 0.003 & 0.045 & 0.022 & 0.011 & $|\cdot|$   & 0.058 & 0.035 & 0.019 \\
      &    & 0.25 & Gaussian & 0.025 & 0.011 & 0.005 & 0.044 & 0.022 & 0.011 & $(\cdot)^2$ & 0.037 & 0.019 & 0.010 \\
      &    &      & Laplace  & 0.012 & 0.005 & 0.003 & 0.045 & 0.022 & 0.011 & $|\cdot|$   & 0.057 & 0.031 & 0.016 \\ \midrule
  2   &  5 & 0.05 & Gaussian & 0.062 & 0.029 & 0.014 & 0.070 & 0.036 & 0.018 & $(\cdot)^2$ & 0.070 & 0.053 & 0.033 \\
      &    &      & Laplace  & 0.031 & 0.014 & 0.007 & 0.073 & 0.036 & 0.018 & $|\cdot|$   & 0.078 & 0.045 & 0.024 \\
      &    & 0.10 & Gaussian & 0.063 & 0.030 & 0.014 & 0.069 & 0.036 & 0.018 & $(\cdot)^2$ & 0.078 & 0.056 & 0.031 \\
      &    &      & Laplace  & 0.031 & 0.015 & 0.007 & 0.073 & 0.036 & 0.018 & $|\cdot|$   & 0.081 & 0.047 & 0.025 \\
      &    & 0.25 & Gaussian & 0.061 & 0.029 & 0.014 & 0.070 & 0.036 & 0.018 & $(\cdot)^2$ & 0.079 & 0.048 & 0.025 \\
      &    &      & Laplace  & 0.030 & 0.014 & 0.007 & 0.072 & 0.036 & 0.018 & $|\cdot|$   & 0.082 & 0.046 & 0.024 \\ \midrule
      & 10 & 0.05 & Gaussian & 0.062 & 0.030 & 0.014 & 0.064 & 0.034 & 0.018 & $(\cdot)^2$ & 0.081 & 0.077 & 0.049 \\
      &    &      & Laplace  & 0.031 & 0.015 & 0.007 & 0.071 & 0.036 & 0.018 & $|\cdot|$   & 0.082 & 0.055 & 0.031 \\
      &    & 0.10 & Gaussian & 0.062 & 0.030 & 0.014 & 0.063 & 0.035 & 0.018 & $(\cdot)^2$ & 0.091 & 0.073 & 0.039 \\
      &    &      & Laplace  & 0.031 & 0.015 & 0.007 & 0.070 & 0.036 & 0.018 & $|\cdot|$   & 0.087 & 0.057 & 0.030 \\
      &    & 0.25 & Gaussian & 0.061 & 0.029 & 0.014 & 0.068 & 0.036 & 0.018 & $(\cdot)^2$ & 0.088 & 0.053 & 0.027 \\
      &    &      & Laplace  & 0.030 & 0.014 & 0.007 & 0.070 & 0.036 & 0.018 & $|\cdot|$   & 0.088 & 0.051 & 0.026 \\ \midrule
  3   &  5 & 0.05 & Gaussian & 0.092 & 0.045 & 0.022 & 0.084 & 0.043 & 0.022 & $(\cdot)^2$ & 0.101 & 0.082 & 0.052 \\
      &    &      & Laplace  & 0.048 & 0.022 & 0.011 & 0.087 & 0.044 & 0.022 & $|\cdot|$   & 0.094 & 0.054 & 0.029 \\
      &    & 0.10 & Gaussian & 0.092 & 0.045 & 0.022 & 0.082 & 0.043 & 0.022 & $(\cdot)^2$ & 0.113 & 0.087 & 0.049 \\
      &    &      & Laplace  & 0.048 & 0.023 & 0.011 & 0.088 & 0.044 & 0.022 & $|\cdot|$   & 0.097 & 0.057 & 0.030 \\
      &    & 0.25 & Gaussian & 0.092 & 0.045 & 0.022 & 0.084 & 0.043 & 0.022 & $(\cdot)^2$ & 0.118 & 0.073 & 0.039 \\
      &    &      & Laplace  & 0.047 & 0.022 & 0.011 & 0.087 & 0.044 & 0.022 & $|\cdot|$   & 0.100 & 0.056 & 0.029 \\ \midrule
      & 10 & 0.05 & Gaussian & 0.085 & 0.044 & 0.022 & 0.074 & 0.040 & 0.022 & $(\cdot)^2$ & 0.104 & 0.115 & 0.077 \\
      &    &      & Laplace  & 0.047 & 0.022 & 0.011 & 0.085 & 0.043 & 0.022 & $|\cdot|$   & 0.097 & 0.067 & 0.038 \\
      &    & 0.10 & Gaussian & 0.086 & 0.045 & 0.022 & 0.073 & 0.042 & 0.022 & $(\cdot)^2$ & 0.123 & 0.110 & 0.062 \\
      &    &      & Laplace  & 0.047 & 0.023 & 0.011 & 0.084 & 0.043 & 0.022 & $|\cdot|$   & 0.103 & 0.069 & 0.037 \\
      &    & 0.25 & Gaussian & 0.090 & 0.045 & 0.022 & 0.080 & 0.043 & 0.022 & $(\cdot)^2$ & 0.129 & 0.081 & 0.042 \\
      &    &      & Laplace  & 0.046 & 0.022 & 0.011 & 0.084 & 0.044 & 0.022 & $|\cdot|$   & 0.106 & 0.061 & 0.032 \\
  \bottomrule
  \end{tabular*}
\end{table*}

We fitted the simulated data using normal and Laplace distributions and performed the restricted estimation under the 
contrast $\mu_1 = \mu_2$. It merits attention that for the Laplace distribution, the covariance matrix adopts the form 
$\cov(\bs{X}) = 4(k+1)\bs{\Sigma}$, and in our case $k=2$. Thus, using Equation \eqref{eq:rhoc-EC}, the Lin's coefficient 
for the bivariate Laplace distribution is given by,
\begin{equation}\label{eq:L1-Laplace}
  \rho_c = \frac{24\sigma_{12}}{12(\sigma_{11} + \sigma_{22}) + (\mu_1 - \mu_2)^2} = \rho\,C_{12}^*,
\end{equation}
where $C_{12}^* = 2(b + b^{-1} + a^2/12)^{-1}$. Furthermore, when $\mu_1 - \mu_2 = 0$, this coefficient has an even simpler 
definition. We developed the estimation of Lin's concordance correlation coefficients under normality and considering the 
Laplace distribution as well as its counterparts based on the $L_1$ distance, i.e., considering Equations \eqref{eq:CCC1} 
and \eqref{eq:CCC1-EC}, respectively. For comparative purposes, we also considered estimating these coefficients using 
$U$-statistics, as introduced by \cite{King:2001}, with the following functions $\varphi(z) = z^2$ and $\varphi(z) = \lvert z\rvert$ 
(see also the Appendix). These estimates are denoted as $\what{\rho}_\varphi$ in Tables \ref{tab:est01} to \ref{tab:est04}.
The asymptotic standard error of the coefficients, $\what{\rho}_c$ and $\what{\rho}_1$, for all simulations is based on 
the results described in Section \ref{sec:inference}. For the standard error of the estimates based on the $U$-statistics 
methodology, refer to \cite{King:2001} and our Appendix.

\begin{table*}[!ht]%
  \caption{Empirical sizes of 5\%, Score, Gradient, Wald and Hotelling's $T^2$ test statistics under Scenarios a), b) and c).\label{tab:test01}}
  \begin{tabular*}{\textwidth}{@{\extracolsep\fill}cclcccccccccccc@{}} \toprule
  Scenario & $m$ & Fitted & \multicolumn{3}{@{}c}{Score} & \multicolumn{3}{@{}c}{Gradient} & \multicolumn{3}{@{}c}{Wald} & \multicolumn{3}{@{}c}{Hotelling's $T^2$} \\ \cmidrule{4-6}\cmidrule{7-9}\cmidrule{10-12}\cmidrule{13-15}
     &   & Model    & 25    & 100   & 400   & 25    & 100   & 400   & 25    & 100   & 400   & 25    & 100   & 400 \\ \midrule
  a) & 1 & Gaussian & 0.054 & 0.054 & 0.051 & 0.054 & 0.054 & 0.051 & 0.075 & 0.060 & 0.053 & 0.069 & 0.059 & 0.052 \\
     &   & Laplace  & 0.020 & 0.014 & 0.016 & 0.076 & 0.076 & 0.070 & 0.191 & 0.189 & 0.182 & 0.069 & 0.059 & 0.052 \\
     & 2 & Gaussian & 0.052 & 0.049 & 0.045 & 0.052 & 0.049 & 0.045 & 0.068 & 0.058 & 0.046 & 0.067 & 0.056 & 0.046 \\
     &   & Laplace  & 0.018 & 0.018 & 0.014 & 0.075 & 0.067 & 0.068 & 0.193 & 0.183 & 0.175 & 0.067 & 0.056 & 0.046 \\
     & 3 & Gaussian & 0.039 & 0.047 & 0.064 & 0.039 & 0.047 & 0.064 & 0.051 & 0.052 & 0.067 & 0.048 & 0.051 & 0.066 \\
     &   & Laplace  & 0.017 & 0.019 & 0.021 & 0.055 & 0.071 & 0.076 & 0.172 & 0.189 & 0.181 & 0.048 & 0.051 & 0.066 \\ \midrule
  b) & 1 & Gaussian & 0.045 & 0.050 & 0.042 & 0.045 & 0.050 & 0.042 & 0.064 & 0.052 & 0.044 & 0.058 & 0.052 & 0.042 \\
     &   & Laplace  & 0.011 & 0.022 & 0.011 & 0.046 & 0.042 & 0.040 & 0.106 & 0.093 & 0.095 & 0.058 & 0.052 & 0.042 \\
     & 2 & Gaussian & 0.052 & 0.047 & 0.041 & 0.052 & 0.047 & 0.041 & 0.065 & 0.050 & 0.041 & 0.064 & 0.050 & 0.041 \\
     &   & Laplace  & 0.016 & 0.012 & 0.012 & 0.052 & 0.040 & 0.044 & 0.127 & 0.098 & 0.103 & 0.064 & 0.050 & 0.041 \\
     & 3 & Gaussian & 0.040 & 0.042 & 0.050 & 0.040 & 0.042 & 0.050 & 0.060 & 0.047 & 0.050 & 0.057 & 0.046 & 0.050 \\
     &   & Laplace  & 0.016 & 0.017 & 0.019 & 0.046 & 0.051 & 0.038 & 0.143 & 0.111 & 0.089 & 0.057 & 0.046 & 0.050 \\ \midrule
  c) & 1 & Gaussian & 0.028 & 0.020 & 0.023 & 0.028 & 0.020 & 0.023 & 0.038 & 0.023 & 0.024 & 0.035 & 0.021 & 0.023 \\
     &   & Laplace  & 0.031 & 0.009 & 0.008 & 0.008 & 0.001 & 0.001 & 0.017 & 0.001 & 0.000 & 0.035 & 0.021 & 0.023 \\
     & 2 & Gaussian & 0.027 & 0.021 & 0.023 & 0.027 & 0.021 & 0.023 & 0.041 & 0.022 & 0.023 & 0.036 & 0.022 & 0.023 \\
     &   & Laplace  & 0.030 & 0.014 & 0.008 & 0.010 & 0.003 & 0.000 & 0.022 & 0.000 & 0.000 & 0.036 & 0.022 & 0.023 \\
     & 3 & Gaussian & 0.011 & 0.016 & 0.021 & 0.011 & 0.016 & 0.021 & 0.023 & 0.021 & 0.021 & 0.018 & 0.020 & 0.021 \\
     &   & Laplace  & 0.039 & 0.010 & 0.004 & 0.016 & 0.002 & 0.000 & 0.017 & 0.001 & 0.000 & 0.018 & 0.020 & 0.021 \\
  \bottomrule
  \end{tabular*}
\end{table*}

\begin{table*}[!ht]%
  \caption{Empirical sizes of 5\%, Score, Gradient, Wald and Hotelling's $T^2$ test statistics under Scenario d).\label{tab:test02}}
  \begin{tabular*}{\textwidth}{@{\extracolsep\fill}ccclcccccccccccc@{}} \toprule
  $m$ & $\eta$ & $\epsilon$ & Fitted & \multicolumn{3}{@{}c}{Score} & \multicolumn{3}{@{}c}{Gradient} & \multicolumn{3}{@{}c}{Wald} & \multicolumn{3}{@{}c}{Hotelling's $T^2$} \\ \cmidrule{5-7}\cmidrule{8-10}\cmidrule{11-13}\cmidrule{14-16}
      &    &      & Model    & 25    & 100   & 400   & 25    & 100   & 400   & 25    & 100   & 400   & 25    & 100   & 400   \\ \midrule
  1   &  5 & 0.05 & Gaussian & 0.036 & 0.032 & 0.048 & 0.036 & 0.032 & 0.048 & 0.062 & 0.039 & 0.049 & 0.051 & 0.038 & 0.049 \\
      &    &      & Laplace  & 0.017 & 0.015 & 0.010 & 0.068 & 0.043 & 0.051 & 0.161 & 0.107 & 0.132 & 0.051 & 0.038 & 0.049 \\
      &    & 0.10 & Gaussian & 0.037 & 0.037 & 0.049 & 0.037 & 0.037 & 0.049 & 0.064 & 0.043 & 0.049 & 0.055 & 0.042 & 0.049 \\
      &    &      & Laplace  & 0.017 & 0.013 & 0.009 & 0.056 & 0.031 & 0.039 & 0.124 & 0.076 & 0.081 & 0.055 & 0.042 & 0.049 \\
      &    & 0.25 & Gaussian & 0.044 & 0.051 & 0.051 & 0.044 & 0.051 & 0.051 & 0.073 & 0.054 & 0.051 & 0.068 & 0.053 & 0.051 \\
      &    &      & Laplace  & 0.012 & 0.013 & 0.009 & 0.032 & 0.016 & 0.019 & 0.080 & 0.036 & 0.042 & 0.068 & 0.053 & 0.051 \\ \midrule
      & 10 & 0.05 & Gaussian & 0.028 & 0.031 & 0.041 & 0.028 & 0.031 & 0.041 & 0.044 & 0.036 & 0.042 & 0.038 & 0.034 & 0.042 \\
      &    &      & Laplace  & 0.019 & 0.012 & 0.009 & 0.055 & 0.036 & 0.026 & 0.117 & 0.072 & 0.064 & 0.038 & 0.034 & 0.042 \\
      &    & 0.10 & Gaussian & 0.027 & 0.037 & 0.042 & 0.027 & 0.037 & 0.042 & 0.046 & 0.043 & 0.043 & 0.039 & 0.041 & 0.042 \\
      &    &      & Laplace  & 0.015 & 0.011 & 0.008 & 0.035 & 0.014 & 0.013 & 0.070 & 0.022 & 0.016 & 0.039 & 0.041 & 0.042 \\
      &    & 0.25 & Gaussian & 0.043 & 0.051 & 0.046 & 0.043 & 0.051 & 0.046 & 0.070 & 0.056 & 0.049 & 0.063 & 0.056 & 0.048 \\
      &    &      & Laplace  & 0.005 & 0.011 & 0.006 & 0.009 & 0.004 & 0.002 & 0.011 & 0.002 & 0.002 & 0.063 & 0.056 & 0.048 \\ \midrule
  2   &  5 & 0.05 & Gaussian & 0.041 & 0.047 & 0.056 & 0.041 & 0.047 & 0.056 & 0.056 & 0.050 & 0.056 & 0.052 & 0.050 & 0.056 \\
      &    &      & Laplace  & 0.014 & 0.014 & 0.019 & 0.058 & 0.061 & 0.058 & 0.145 & 0.133 & 0.125 & 0.052 & 0.050 & 0.056 \\
      &    & 0.10 & Gaussian & 0.035 & 0.047 & 0.052 & 0.035 & 0.047 & 0.052 & 0.051 & 0.054 & 0.052 & 0.042 & 0.052 & 0.052 \\
      &    &      & Laplace  & 0.013 & 0.015 & 0.017 & 0.050 & 0.044 & 0.046 & 0.112 & 0.090 & 0.092 & 0.042 & 0.052 & 0.052 \\
      &    & 0.25 & Gaussian & 0.032 & 0.049 & 0.056 & 0.032 & 0.049 & 0.056 & 0.050 & 0.057 & 0.059 & 0.047 & 0.056 & 0.057 \\
      &    &      & Laplace  & 0.012 & 0.014 & 0.016 & 0.031 & 0.020 & 0.035 & 0.059 & 0.039 & 0.050 & 0.047 & 0.056 & 0.057 \\ \midrule
      & 10 & 0.05 & Gaussian & 0.028 & 0.030 & 0.044 & 0.028 & 0.030 & 0.044 & 0.042 & 0.039 & 0.046 & 0.036 & 0.035 & 0.046 \\
      &    &      & Laplace  & 0.016 & 0.013 & 0.017 & 0.044 & 0.036 & 0.045 & 0.116 & 0.074 & 0.068 & 0.036 & 0.035 & 0.046 \\
      &    & 0.10 & Gaussian & 0.026 & 0.037 & 0.047 & 0.026 & 0.037 & 0.047 & 0.039 & 0.042 & 0.047 & 0.035 & 0.042 & 0.047 \\
      &    &      & Laplace  & 0.013 & 0.009 & 0.011 & 0.028 & 0.014 & 0.021 & 0.070 & 0.028 & 0.025 & 0.035 & 0.042 & 0.047 \\
      &    & 0.25 & Gaussian & 0.024 & 0.052 & 0.053 & 0.024 & 0.052 & 0.053 & 0.046 & 0.053 & 0.050 & 0.042 & 0.052 & 0.053 \\
      &    &      & Laplace  & 0.008 & 0.008 & 0.007 & 0.007 & 0.004 & 0.002 & 0.008 & 0.004 & 0.002 & 0.042 & 0.052 & 0.053 \\ \midrule
  3   &  5 & 0.05 & Gaussian & 0.030 & 0.040 & 0.046 & 0.030 & 0.040 & 0.046 & 0.046 & 0.045 & 0.046 & 0.043 & 0.043 & 0.046 \\
      &    &      & Laplace  & 0.011 & 0.010 & 0.011 & 0.049 & 0.038 & 0.053 & 0.148 & 0.113 & 0.121 & 0.043 & 0.043 & 0.046 \\
      &    & 0.10 & Gaussian & 0.023 & 0.041 & 0.045 & 0.023 & 0.041 & 0.045 & 0.040 & 0.044 & 0.048 & 0.036 & 0.044 & 0.048 \\
      &    &      & Laplace  & 0.010 & 0.010 & 0.009 & 0.038 & 0.031 & 0.033 & 0.102 & 0.076 & 0.090 & 0.036 & 0.044 & 0.048 \\
      &    & 0.25 & Gaussian & 0.030 & 0.035 & 0.058 & 0.030 & 0.035 & 0.058 & 0.044 & 0.038 & 0.058 & 0.041 & 0.037 & 0.058 \\
      &    &      & Laplace  & 0.010 & 0.007 & 0.008 & 0.015 & 0.012 & 0.015 & 0.044 & 0.028 & 0.035 & 0.041 & 0.037 & 0.058 \\ \midrule
      & 10 & 0.05 & Gaussian & 0.021 & 0.037 & 0.037 & 0.021 & 0.037 & 0.037 & 0.040 & 0.039 & 0.038 & 0.032 & 0.038 & 0.038 \\
      &    &      & Laplace  & 0.011 & 0.010 & 0.009 & 0.040 & 0.025 & 0.026 & 0.107 & 0.066 & 0.065 & 0.032 & 0.038 & 0.038 \\
      &    & 0.10 & Gaussian & 0.018 & 0.033 & 0.052 & 0.018 & 0.033 & 0.052 & 0.031 & 0.039 & 0.054 & 0.028 & 0.038 & 0.054 \\
      &    &      & Laplace  & 0.010 & 0.007 & 0.006 & 0.024 & 0.014 & 0.010 & 0.054 & 0.027 & 0.019 & 0.028 & 0.038 & 0.054 \\
      &    & 0.25 & Gaussian & 0.032 & 0.036 & 0.061 & 0.032 & 0.036 & 0.061 & 0.050 & 0.044 & 0.062 & 0.046 & 0.043 & 0.061 \\
      &    &      & Laplace  & 0.007 & 0.003 & 0.002 & 0.004 & 0.002 & 0.001 & 0.008 & 0.001 & 0.000 & 0.046 & 0.043 & 0.061 \\
  \bottomrule
  \end{tabular*}
\end{table*}

The estimation results from 1000 runs are reported in Tables \ref{tab:est01} and \ref{tab:est03}. When the data follow normal 
or Laplace distributions (Scenarios a and b), the estimated coefficients $\what{\rho}_c$ and $\what{\rho}_1$ for each fitted 
model perform similarly and are close to the reference values. The situation changes for extreme contamination levels, such 
as that offered by the Cauchy distribution, which has extremely heavy tails. We should emphasize that the estimate of Lin's 
coefficient of concordance, $\what{\rho}_c = \what{\rho}\,\what{C}{}_{12}^*$, given in \eqref{eq:L1-Laplace} performs better 
than its counterpart under the normal distribution, with consistently lower standard errors (see Table \ref{tab:est02}). A 
similar situation occurs for the contaminated normal distribution when $m = 2$ or 3 with moderate contamination percentages, 
i.e., $\epsilon = 10\%$ or 25\%, regardless of the level of the variance inflation factor. As expected, our version based on 
$L_1$ distance is not strongly affected by outliers. Both the version obtained under normality and the version calculated under 
a Laplace distribution behave similarly, with the version given by Equation \eqref{eq:CCC1-EC} performing slightly better, even 
for small samples, with very similar standard errors. Interestingly, in the presence of high levels of contamination, such as 
data generated under the Cauchy or the contaminated normal distributions with $\epsilon = 10\%$ or 25\%, the estimation offered 
by the $U$-statistics method using $\varphi(z) = \lvert z\rvert$ has a tendency to overestimate the coefficient $\rho_1$. Additionally, 
for all simulations where there is some level of contamination, the estimated coefficients $\what{\rho}_{\varphi}$ present larger standard 
errors than their counterparts based on the Laplace distribution, an aspect that is expected due to the distribution-free nature 
of the $U$-statistics procedure.

Tables \ref{tab:test01} and \ref{tab:test02} report the empirical sizes of different test statistics described in 
Section \ref{sec:inference}. For low levels of contamination the results are consistently close to the nominal size, 
while the likelihood-based tests tend to be more conservative for high levels of contamination. It is worth emphasizing the strong performance of the Hotelling $T^2$
  statistic, except in cases of severe contamination from the Cauchy distribution. As expected, the generated data satisfies the assumption of equality of means. This allows us to apply 
the result displayed in Equation \ref{eq:asymp-rho1}.

\subsection{A real data application: Transient sleep disorder}

\begin{figure}[!ht]
  \vskip -1.75em
  \centering
  \subfigure[]{
    \includegraphics[width = 0.42\linewidth]{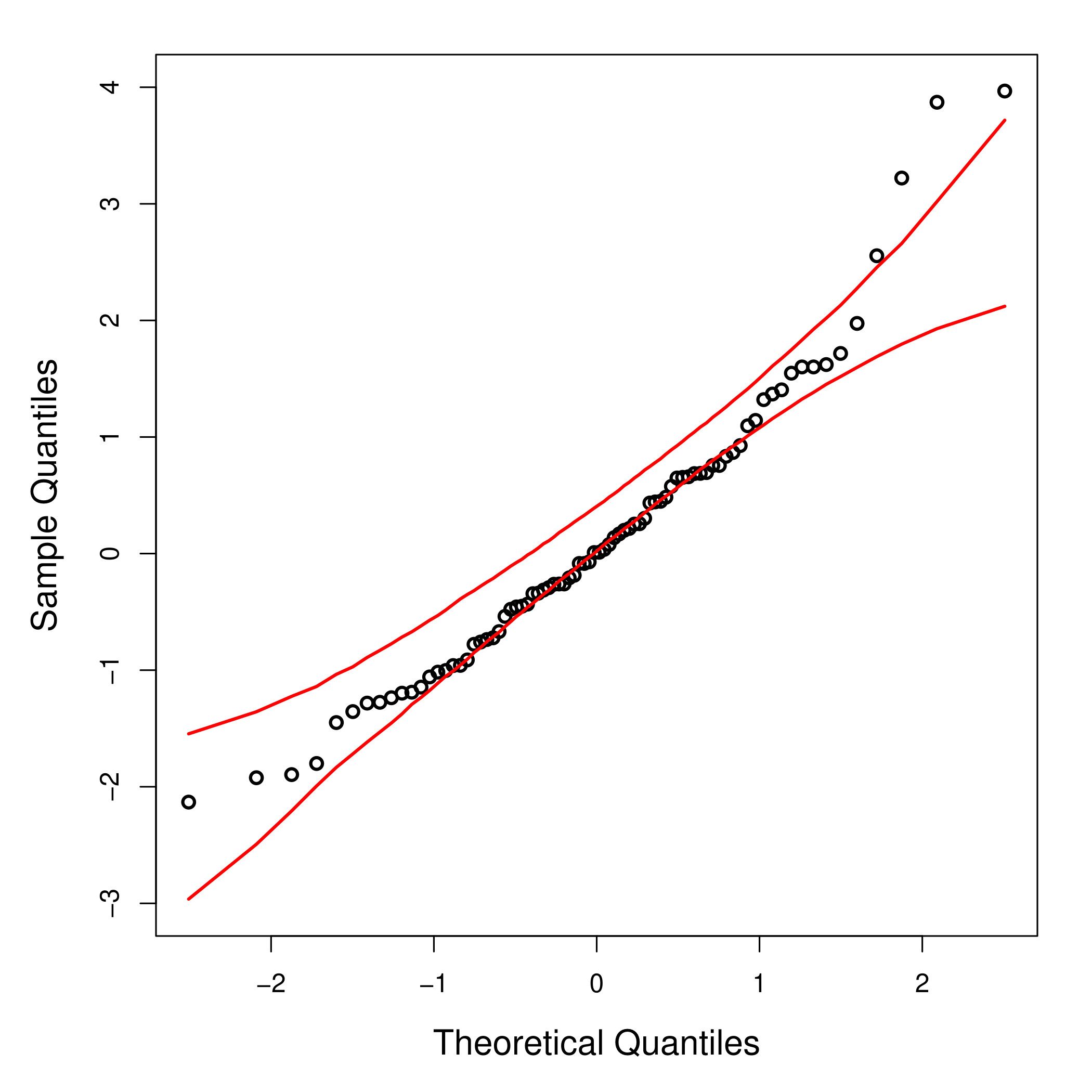}
  }
  \subfigure[]{
    \includegraphics[width = 0.45\linewidth]{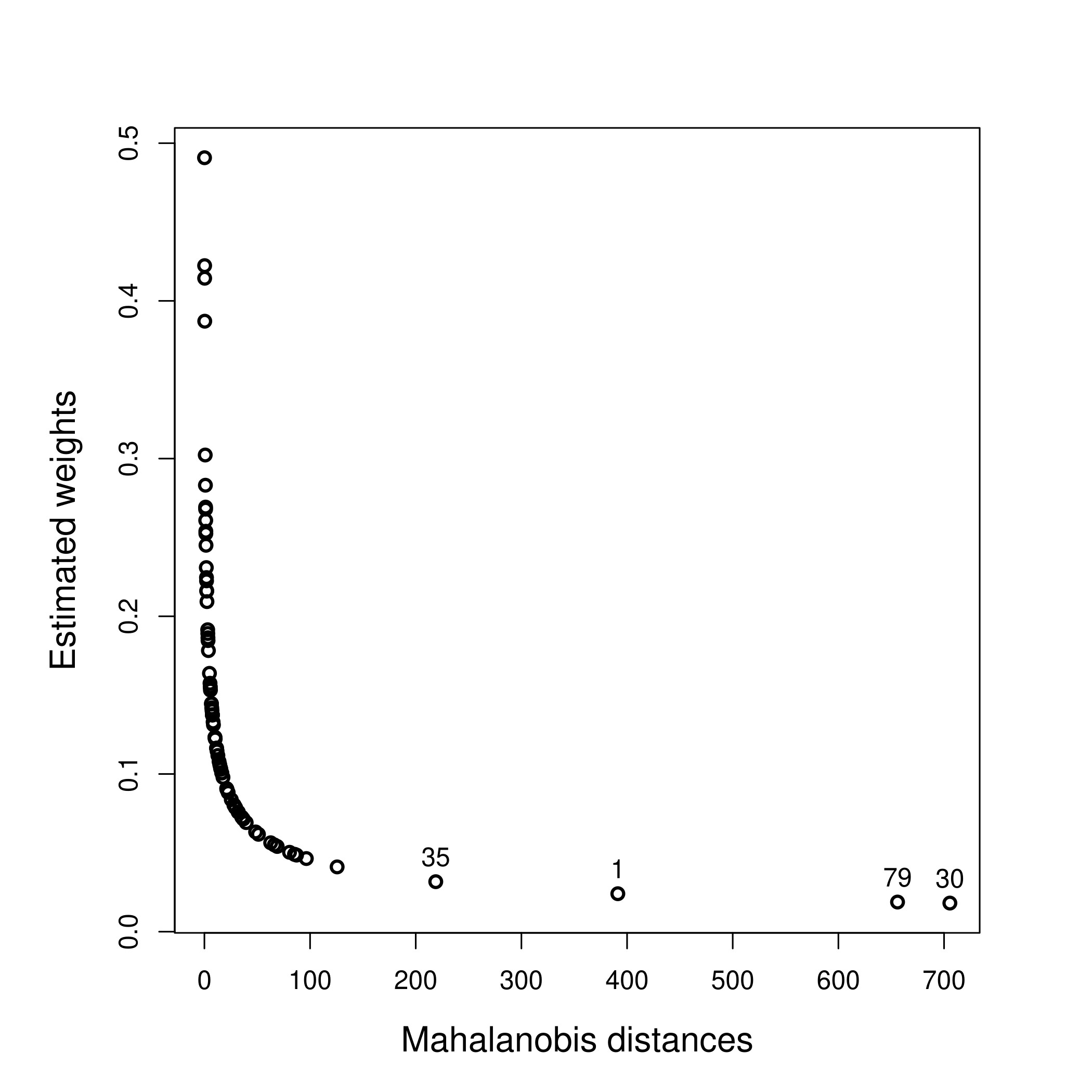}
  }
  \caption{(a) QQ-plot of the transformed distances, and (b) Estimated weights against Mahalanobis distances, 
  fitted model under the multivariate Laplace distribution.}\label{fig:fit}
\end{figure}

Next, we consider the dataset for the study of insomnia problems described in Section \ref{sec:intro}. This dataset is included 
in the R package \textsf{MVT}\cite{MVT} available from CRAN repository. Interestingly, the study by \cite{Osorio:2023} reports that the 
normality assumption is not supported by the data (see also Figure \ref{fig:intro}\,(b)). In particular, by introducing a perturbation 
to the Gaussian model, they found that observations 30 and 79 are strongly influential, a conclusion that is consistent with that 
obtained in the diagnostic analysis reported in \cite{Leal:2019}. The main conclusion of the latter study is the extreme sensitivity 
of Lin's concordance correlation coefficient to outlying observations. Whereas the QQ-plot displayed in Figure \ref{fig:fit}\,(a) 
reveals that the Laplace distribution can be used to reasonably fit this type of data.  The estimation 
procedure effectively downweights those observations identified as outliers (see Figure \ref{fig:fit}\,(b) and compare with the 
influence analysis in \cite{Leal:2019, Osorio:2023}). 

Tables \ref{tab:1} and \ref{tab:2} present the fit using the normal and multivariate Laplace model and under the constraint $\mu_1 
= \mu_2$, respectively. Note that in those tables the covariance matrix is reported, which for the case of the bivariate Laplace 
distribution is given by $\cov(\bs{X}) = 12\bs{\Sigma}$. It is noteworthy that although the Laplace distribution provides a better 
fit than the multivariate normal distribution, Table \ref{tab:3} shows that in both cases the hypothesis $H_0: \mu_1=\mu_2$ is 
rejected for each of the test statistics used. The same conclusion is obtained by using the generalized $T^2$ statistic. Thus the 
assumption required by Propositions \ref{prop:3} and \ref{prop:4} is not supported by the data. Table \ref{tab:4} presents the 
estimates of Lin's and $L_1$ concordance coefficients considering the normal and multivariate Laplace distributions and under 
the assumption of equality of means between the measurement instruments. We calculated the standard errors in cases where the 
assumption $\mu_1 = \mu_2$ is not satisfied using bootstrap, which is implemented in the \textsf{L1pack} package. For comparative 
purposes, we have calculated the concordance coefficient using $U$-statistics with $\varphi(z) = \lvert z\rvert$, obtaining 
$\what{\rho}_\varphi = 0.6577$ and $\SE(\what{\rho}_\varphi) = 0.0598$ (details are relegated to Appendix).

Interestingly, the level of agreement between the instruments varies greatly depending on the distributional assumption and the 
coefficient used. Moreover,  the variance of the coefficient proposed in this study is much lower than its counterpart 
based on Lin's coefficient. Additionally, the use of a coefficient with poor robustness characteristics can distort the level of 
agreement in the presence of outlying observations, even when using a distribution that fits the data appropriately.

\begin{table*}[!ht]%
  \caption{Parameter estimates under gaussian and Laplace assumptions.\label{tab:1}}
  \begin{tabular*}{\textwidth}{@{\extracolsep\fill}lcccccc@{}} \toprule
    & \multicolumn{3}{@{}c}{\textbf{Gaussian}} & \multicolumn{3}{@{}c}{\textbf{Laplace}} \\ \cmidrule{2-4}\cmidrule{5-7}
  \textbf{Variable} & \textbf{mean} & \multicolumn{2}{@{}c}{\textbf{covariance matrix}} & \textbf{mean} & \multicolumn{2}{@{}c}{\textbf{covariance matrix}} \\ \midrule
  manual         & 2.5539 & 0.7617 &        & 2.5691 & 0.7132 &        \\
  automated      & 2.3090 & 0.6942 & 1.2369 & 2.4339 & 0.6928 & 0.9109 \\ \midrule
  log-likelihood & \multicolumn{3}{@{}l}{-200.8901} & \multicolumn{3}{@{}l}{-179.9035} \\
  \bottomrule
  \end{tabular*}
\end{table*}

\begin{table*}[!ht]%
  \caption{Restricted parameter estimates under the constrain $\mu_1 = \mu_2$ for gaussian and Laplace distributions.\label{tab:2}}
  \begin{tabular*}{\textwidth}{@{\extracolsep\fill}lcccccc@{}} \toprule
    & \multicolumn{3}{@{}c}{\textbf{Gaussian}} & \multicolumn{3}{@{}c}{\textbf{Laplace}} \\ \cmidrule{2-4}\cmidrule{5-7}
  \textbf{Variable} & \textbf{mean} & \multicolumn{2}{@{}c}{\textbf{covariance matrix}} & \textbf{mean} & \multicolumn{2}{@{}c}{\textbf{covariance matrix}} \\ \midrule
  manual         & 2.5268 & 0.7624 &        & 2.5299 & 0.7254 &        \\
  automated      & 2.5268 & 0.6883 & 1.2844 & 2.5299 & 0.6952 & 0.9245 \\ \midrule
  log-likelihood & \multicolumn{3}{@{}l}{-204.7342} & \multicolumn{3}{@{}l}{-183.7141} \\
  \bottomrule
  \end{tabular*}
\end{table*}

\begin{table*}[!ht]%
  \caption{Test $H_0:\mu_1 = \mu_2$ under gaussian and Laplace assumptions.\label{tab:3}}
  \begin{tabular*}{\textwidth}{@{\extracolsep\fill}lccrc@{}} \toprule
    & \multicolumn{2}{@{}c}{\textbf{Gaussian}} & \multicolumn{2}{@{}c}{\textbf{Laplace}} \\ \cmidrule{2-3}\cmidrule{4-5}
  \textbf{Test} & \textbf{statistic} & \textbf{\emph{p}-value} & \textbf{statistic} & \textbf{\emph{p}-value} \\ \midrule
  Wald     & 8.0600 & 0.0045 &  9.4267 & 0.0021 \\
  score    & 7.3387 & 0.0067 &  7.4012 & 0.0065 \\
  gradient & 7.3387 & 0.0067 & 16.3815 & 0.0001 \\
  LRT      & 7.6881 & 0.0056 &  7.6211 & 0.0058 \\
  \bottomrule
  \end{tabular*}
  \begin{tablenotes}
    \item $T^2 = 7.9617$, $p\textrm{-value} = 0.0048$.
  \end{tablenotes}
\end{table*}

\begin{table*}[!ht]%
  \caption{Estimates of agreement coefficients (standard errors in parenthesis) under gaussian and Laplace assumptions.\label{tab:4}}
  \begin{tabular*}{0.7\textwidth}{@{\extracolsep\fill}lcc@{}} \toprule
  \textbf{Coefficient} & \textbf{Gaussian} & \textbf{Laplace} \\ \midrule
  Lin's                       & 0.6744 (0.0563) & 0.8436 (0.0643) \\
  Lin's under $\mu_1 = \mu_2$ & 0.6726 (0.0575) & 0.8428 (0.0660) \\
  $L_1$                       & 0.4291 (0.0930) & 0.5855 (0.0804) \\
  $L_1$ under $\mu_1 = \mu_2$ & 0.4278 (0.0502) & 0.6035 (0.0398) \\
  \bottomrule
  \end{tabular*}
\end{table*}

\section{Discussion}

In Section \ref{sec:intro}, we emphasized that the proposed methodology is guided by two key principles: the development of a robust concordance coefficient that performs well in the presence of influential observations or outliers, and the avoidance of additional parameters that require fixing or estimation, as is common in classical robust approaches. The results presented in this paper demonstrate that both of these objectives have been successfully achieved.

We have developed a version of the $L_1$ distance-based concordance correlation coefficient, and examined its properties for the broad class of elliptically contoured distributions. Two side results are of considerable interest. First, when the means of the measuring instruments are equal, our coefficient is a simple transformation of Lin's concordance coefficient, which greatly simplifies the study of its properties, in particular its asymptotic distribution. Additionally, it allows us to extend our proposal by considering a $L_p$ distance. Second, under specific conditions, the choice of $p = 1$ minimizes the asymptotic variance of our coefficient of concordance.

We focused on the Laplace distribution as it corresponds to a member of the elliptical class, which is often suggested to lead to estimators that minimize the $L_1$ distance. To guide the practitioner as to whether the data satisfy the assumption of equality of means we have made available likelihood-based tests as well as Hotelling's $T^2$ statistic which only requires second moments and performed very well in our numerical experiments. Additionally, Section \ref{sec:gof} presents a graphical device that enables the easy verification of the distributional assumption.

This work represents a preliminary treatment of concordance measures under an 
$L_1$ framework, opening multiple avenues for future investigation. For instance, one potential direction is to extend these coefficients to the agreement of two digital images. A nonparametric approach could be particularly advantageous, as it avoids the need for strong distributional assumptions—an important consideration given that image textures often exhibit complex structures that can only be effectively modeled using parametric methods within a local framework \cite{Vallejos:2020}. A natural approach to addressing concordance in images is to employ a kernel smoothing estimator that incorporates a discrepancy measure between pixels and a kernel function to appropriately weight neighboring values (e.g., \cite{Datta:2016}).

\bmsection*{Acknowledgments}

This work has been supported by ANID through Fondecyt grant 1230012. Ronny Vallejos also has been partially supported 
by the AC3E, UTFSM, under grant AFB240002. Felipe Osorio's research benefited from financial support provided by the Universidad 
T\'{e}cnica Federico Santa Mar\'{i}a via the grant PI\_LIR\_24\_02. The authors express their gratitude to Jonathan Acosta 
for engaging and insightful discussions and for revising a preliminary version of this manuscript.

\bmsection*{Data Availability Statement}

The replication files related to this article are available online at \url{https://github.com/faosorios/L1CCC}


\bmsection*{Conflict of interest}

The authors declare no potential conflict of interests.

\bmsection*{ORCID}

Ronny Vallejos: \url{https://orcid.org/0000-0001-5519-0946} \\
Felipe Osorio: \url{https://orcid.org/0000-0002-4675-5201} 

\appendix

\bmsubsection*{A scaled $\bs{\rho_1}$ coefficient}

Consider the following result established by \cite{King:2001}. Assume that $(X_1,X_2)^\top$ 
follows a bivariate normal distribution with $\bs{\mu} = (\mu_1,\mu_2)^\top$, and 
\[
  \bs{\Sigma} = \begin{pmatrix}
    \sigma_{11} & \sigma_{12} \\
    \sigma_{21} & \sigma_{22}
  \end{pmatrix},
\]
and suppose that $\varphi(z)$ is a non-decreasing convex function such that for all $z\geq 0$, 
$\varphi(0)=0$ and $\varphi(z) = \varphi(-z)$. Then, the coefficient
\[
  \rho_\varphi = \frac{\E\{\varphi(X_1-X_2):\sigma_{12} = 0\} - \E\{\varphi(X_1+X_2):\sigma_{12}=0\} 
  - \big[\E\{\varphi(X_1-X_2)\} - \E\{\varphi(X_1+X_2)\}\big]}{\E\{\varphi(X_1-X_2):\sigma_{12}=0\} 
  - \E\{\varphi(X_1+X_2):\sigma_{12}=0\} + \half\big[\E\{\varphi(2X_1)\} + \E\{\varphi(2X_2)\}\big]},
\]
belongs to the interval $[-1,1]$. It is straightforward to see that Lin's coefficient given in \eqref{eq:Lin} 
is obtained for $\varphi(z) = z^2$. Similarly, an $L_1$ coefficient like \eqref{eq:CCC1} can be derived using 
$\varphi(z) = \lvert z\rvert$. This coefficient will be a scaled version of \eqref{eq:CCC1} to the interval $[-1,1]$.

\bmsubsection*{Proofs of main results}

\begin{proof}[Proof of Proposition~{\rm\ref{prop:1}}]
  Note that $X_1 - X_2\sim \mathsf{EC}_1(\gamma,\tau^2;\psi)$. Also notice that $Z = \gamma + \tau W$, 
  where $\gamma = \mu_1-\mu_2$, $\tau^2 = \sigma_{11}+\sigma_{22}-2\sigma_{12}$ and $W\sim\mathsf{EC}_1(0,1;\psi)$, 
  thus $\lvert Z\rvert$ follows a folded elliptical contoured distribution and $\lvert Z\rvert = \lvert X_1 - X_2\rvert$. 
  Then,
  \begin{align*}
    \E(\lvert Z\rvert) & = \E(\lvert\gamma + \tau W\rvert) = \E[(\gamma + \tau W)I\{\gamma + \tau W\geq 0\}]
    + \E[(-\gamma - \tau W)I\{\gamma + \tau W < 0\}] \\[.25em]
    & = \E(\gamma + \tau W) - 2\E[(\gamma + \tau W)I\{W < - \gamma/\tau\}] = \gamma - 2\E[\gamma\,
    I\{W < -\gamma/\tau\}] - 2\tau\E[W\,I\{W < -\gamma/\tau\}] \\
    & = \gamma - 2\gamma\Pr(W \leq -\gamma/\tau) - 2\tau\E[W\,I\{W < -\gamma/\tau\}] = \gamma - 2\gamma
    C_g\int_{-\infty}^{-\gamma/\tau} g(r^2)\rd r - 2\tau C_g\int_{-\infty}^{-\gamma/\tau} r g(r^2)\rd r \\
    & = \gamma\Big(1 - 2C_g\int_{-\infty}^{-\gamma/\tau} g(r^2)\rd r\Big) - 2\tau C_g
    \int_{-\infty}^{-\gamma/\tau} r g(r^2)\rd r.
  \end{align*}
  Thus, this yields \eqref{eq:CCC1-EC} and the proposition is verified.
\end{proof}

\begin{proof}[Proof of Proposition~{\rm\ref{prop:2}}]
  Because of \eqref{eq:density},
  \begin{align*}
    \E(\lvert Z\rvert^p) & =\frac{C_g}{\sqrt{\sigma_{11}+\sigma_{22}-2\sigma_{12}}} \int_{-\infty}^\infty \lvert z\rvert^p
    g\Big(\frac{z^2}{\sigma_{11}+\sigma_{22}-2\sigma_{12}}\Big)\rd z \\
    & = \frac{2C_g}{\sqrt{\sigma_{11}+\sigma_{22}-2\sigma_{12}}} \int_0^\infty z^p g\Big(\frac{z^2}{\sigma_{11}+\sigma_{22}
    -2\sigma_{12}}\Big)\rd z \\
    & = (\sigma_{11} + \sigma_{22} - 2\sigma_{12})^{p/2} G_p(t),
  \end{align*}
  where $G_p(t) = \displaystyle \int_0^\infty t^{(p-1)/2}g(t)\rd t$. Thus, 
  \[
    \rho_p = 1 - \frac{\E(\lvert X_1-X_2\rvert^p)}{\E(\lvert X_1-X_2\rvert^p: \sigma_{12}=0)} 
    = 1 - \Big(\frac{\sigma_{11} + \sigma_{22} - 2\sigma_{12}}{\sigma_{11}+\sigma_{22}}\Big)^{p/2}.
  \]
\end{proof}

\begin{proof}[Proof of Proposition~{\rm\ref{prop:3}}]
  Since $\sigma_{12} = \rho\sqrt{\sigma_{11}\sigma_{22}}$ and $\mu_1=\mu_2$,
  \[
    v^2= \frac{1}{n-2}\Big\{\frac{(1-\rho^2)\rho_c^2}{(1-\rho_c^2)\rho^2} + \frac{4\rho_c^3(1-\rho_c)u^2}{\rho(1-\rho_c^2)^2}
    - \frac{2\rho_c^2u^4}{\rho^2(1-\rho_c^2)^2}\Big\} = \frac{(1-\rho^2)\rho_c^2}{(n-2)(1-\rho_c^2)\rho^2}.
  \]
   A straightforward calculation shows that
   \begin{align*}
    \var(\what{\rho}_p) & = \frac{1}{4}\rho v^2(1-\rho_c^2)^2\big\{1-\rho_c)^{p/2-1}\big\}^2 \\
    & = \frac{p^2 \left(1-\rho^2\right)\sigma _{11}\sigma_{22}\big(2\rho\sqrt{\sigma_{11}\sigma_{22}}
    + \sigma_{11} + \sigma_{22}\big)\big(1 - 2\rho\sqrt{\sigma_{11}\sigma_{22}}/(\sigma_{11} + 
    \sigma_{22})\big)^{p-1}}{(n-2)(\sigma_{11} + \sigma_{22})^3}.
  \end{align*}
  If $\rho<0$, then $1 - 2\rho\sqrt{\sigma_{11}\sigma_{22}}/(\sigma_{11} + 
  \sigma_{22})>1$ and the function $p\mapsto p^2\big(1 - 2\rho\sqrt{\sigma_{11}\sigma_{22}}/(\sigma_{11} + 
  \sigma_{22})\big)^{p-1}$ is strictly increasing in $p$ for $p>0$. Therefore, the minimum of $\var(\what{\rho}_p)$ occurs at $p=1$.
\end{proof}

\bmsubsection*{ML estimation under equality of location parameters}

Consider $\bs{x}_1,...,\bs{x}_n$ as a random sample from a population following a $\mathsf{Laplace}_k(\bs{\mu},\bs{\Sigma})$ 
distribution, and consider the constraint,
\[
  \mu_1 = \mu_2 = \cdots = \mu_k = \lambda,
\]
that is, we can write $\bs{\mu} = \lambda\bs{1}_k$ with $\lambda\in\Rset$. Assume the hierarchical 
model,
\[
  \bs{X}_i\mid\omega_i \stackrel{\sf ind}{\sim} \mathsf{N}_k(\lambda\bs{1}_k,\bs{\Sigma}),
\]
and $\omega_i$ follows the distribution given by the density defined in Equation \eqref{eq:mixture}.
Thus, an EM algorithm\cite{Meng:1993} to perform the estimation subject to the constraint of equality 
among the location parameters is summarized in the following steps,
\begin{itemize}
  \item \textbf{E-step:} Using a current estimate for $\bs{\theta} = (\lambda,\bs{\sigma}^\top)^\top$, 
  the conditional expectation of the log-likelihood function for complete data assumes the form,
  \[
    Q(\bs{\theta};\bs{\theta}^{(r)}) = -\frac{n}{2}\log\lvert\bs{\Sigma}\rvert - \frac{1}{2}
    \sum_{i=1}^n \omega_i^{(r)}(\bs{x}_i - \lambda\bs{1}_k)^\top\bs{\Sigma}^{-1}(\bs{x}_i - \lambda\bs{1}_k),
  \]
  where the weights $\omega_i^{(r)}$ are defined in Equation \eqref{eq:E-step}, although they must be 
  evaluated at the distances,
  \[
    D_i(\bs{\theta}^{(r)}) = \big[(\bs{x}_i - \lambda\bs{1}_k)^\top\{\bs{\Sigma}^{(r)}\}^{-1}
    (\bs{x}_i - \lambda\bs{1}_k)\big]^{1/2}, \qquad i=1,\dots,n.
  \]

  \item \textbf{CM-step 1:} Update $\lambda^{(r+1)}$, as:
  \[
    \lambda^{(r+1)} = \frac{\bs{1}_k^\top\{\bs{\Sigma}^{(r)}\}^{-1}\bs{\mu}^{(r+1)}}
    {\bs{1}_k^\top\{\bs{\Sigma}^{(r)}\}^{-1}\bs{1}_k},
  \]
  where $\bs{\mu}^{(r+1)}$ is given in \eqref{eq:M-center}.

  \item \textbf{CM-step 2:} Update $\bs{\Sigma}^{(r+1)}$ by maximizing $Q(\bs{\theta};\bs{\theta}^{(r)})$ 
  with respect to $\bs{\sigma}$, leading 
  \[
    \bs{\Sigma}^{(r+1)} = \frac{1}{n}\,\sum_{i=1}^n \omega_i^{(r)}(\bs{x}_i - \lambda^{(r+1)}\bs{1}_k)
    (\bs{x}_i - \lambda^{(r+1)}\bs{1}_k)^\top.
  \]
\end{itemize}
Thus, the procedure iterates between the E-step and the CM-steps defined above until convergence is reached.

\bmsubsection*{Asymptotic distribution of the $\bs{U}$-statistic based $\bs{\what{\rho}_1}$ coefficient}

Consider $(X_{11},X_{21})^\top,\dots,(X_{1n},X_{2n})^\top$ a random sample from a bivariate distribution with marginal 
and joint cumulative distribution functions $F_1$, $F_2$, and $F_{12}$, respectively. Following \cite{King:2001} we can 
notice that the estimator of the coefficient $\rho_1$ can be written as,
\[
  \what{\rho}_1 = 1 - \frac{\sum_{i=1}^n \lvert X_{1i} - X_{2i}\rvert/n}{\sum_{i=1}^n\sum_{j=1}^n \lvert X_{1i} - X_{2j}\rvert/n^2} 
  = \frac{\sum_{i=1}^n\sum_{j=1}^n \lvert X_{1i} - X_{2j}\rvert/n - \sum_{i=1}^n \lvert X_{1i} - X_{2i}\rvert}{\sum_{i=1}^n\sum_{j=1}^n \lvert X_{1i} - X_{2j}\rvert/n}
\]
Define the following $U$-statistics,
\[
  U_1 = \frac{1}{n(n-1)}\sum_{i\neq j}^n \half\{\lvert X_{1i} - X_{2i}\rvert + \lvert X_{1j} - X_{2j}\rvert\}, \qquad
  U_2 = \frac{1}{n(n-1)}\sum_{i\neq j}^n \half\{\lvert X_{1i} - X_{2j}\rvert + \lvert X_{1j} - X_{2i}\rvert\},
\]
which leads to
\[
  \what{\rho}_1 = \frac{U_1 + (n-1)U_2 - nU_1}{U_1 + (n-1)U_2} = \frac{(n-1)(U_2 - U_1)}{U_1 + (n-1)U_2}
\]
That is, we can write $\what{\rho}_1$ as the ratio between $H = (n-1)(U_2 - U_1)$ and $G = U_1 + (n-1)U_2$, where
the statistics $U_1$ and $U_1 + (n-1)U_2$ are unbiased estimates of $\E(\lvert X_1 - X_2\rvert)$ and $n\E(\lvert X_1 
- X_2\rvert : \sigma_{12} = 0)$, respectively. By using results available in \cite{King:2001}, it follows that 
$\what{\rho}_1$ is asymptotically normal with mean $\rho_1 = \E(H)/\E(G)$ and variance,
\[
  \var(\what{\rho}_1) = \what{\rho}{}_1^2\Big\{\frac{\var(H)}{H^2} + \frac{\var(G)}{G^2} 
  - \frac{2\cov(H,G)}{HG}\Big\},
\]
where,
\begin{align*}
  \var(H) & = (n-1)^2\{\var(U_1) + \var(U_2) - 2\cov(U_1,U_2)\}, \\
  \var(G) & = (n-1)^2\var(U_2) + \var(U_1) + 2(n-1)\cov(U_1,U_2), \\
  \cov(H,G) & = (n-1)\{(n-1)\var(U_2) - \var(U_1) - (n-2)\cov(U_1,U_2)\}. 
\end{align*}
Additional details on the estimation of the asymptotic variance, $\var(\what{\rho}_1)$, are presented in Appendix II 
of \cite{King:2001}.


\end{document}